\documentclass[aip,jcp,
amsmath,amssymb,
reprint,
]{revtex4-2}
\usepackage[T1]{fontenc}
\usepackage[utf8]{inputenc}
\usepackage{hyperref}
\usepackage{graphicx}
\usepackage{dcolumn}
\usepackage{bm}
\usepackage{xcolor}
\usepackage[normalem]{ulem}
\usepackage{enumitem}
\usepackage{comment}
\usepackage{multirow}
\usepackage{natbib}

\begin{document}

\title{Computational Methods toward Ultrastable Glasses}

\author{Fabio Leoni}
\affiliation{Department of Physics, Sapienza University of Rome, Piazzale Aldo Moro 2, 00185 Roma, Italy}
\author{Misaki Ozawa}
\affiliation{Univ. Grenoble Alpes, CNRS, LIPhy, 38000 Grenoble, France}
\author{John Russo}
\affiliation{Department of Physics, Sapienza University of Rome, Piazzale Aldo Moro 2, 00185 Roma, Italy}
\author{Taiki Yanagishima}
\affiliation{Department of Physics, Graduate School of Science, Kyoto University, Kyoto 606-8502, Japan}
\affiliation{Department of Physics, Tokyo Metropolitan University, Tokyo 192-0397, Japan}
\author{Andrea Ninarello}
 \email{andrea.ninarello@cnr.it}
\affiliation{CNR Institute of Complex Systems, Uos Sapienza, Piazzale Aldo Moro 2, 00185, Roma, Italy}
\affiliation{Department of Physics, Sapienza University of Rome, Piazzale Aldo Moro 2, 00185 Roma, Italy}

\begin{abstract}
Ultrastable glasses, amorphous solids with exceptionally low-energy states and enhanced kinetic, thermodynamic and mechanical stability, have long been a subject of intense experimental interest. Over the past decade, their computational realization has emerged as a major goal in condensed matter physics, as numerical methods can exploit unphysical moves to access deeply supercooled and nonequilibrium glassy states far beyond the reach of conventional cooling protocols, thereby providing key insights into the nature of the glass transition and amorphous states and enabling the design of mechanically robust glassy materials.
In this review, we outline the key steps underlying the most effective algorithms developed across the field.
For each approach, we discuss its efficiency, limitations, and physical interpretation. We finally present a comparative analysis of the stability achieved across these methods, with the aim of equipping both newcomers and experts with an intuitive and comprehensive understanding of the field's current state and the opportunities it presents.

\end{abstract}

\maketitle

\tableofcontents

\section{Introduction}
\label{sec:intro}

Glasses have accompanied human technology since antiquity and today underpin applications ranging from optics to electronics. Despite their ubiquity, they remain conceptually intriguing because rigidity emerges without long-range order and the resulting materials are inherently out of equilibrium~\cite{ediger1996, berthier2011}. A particularly active frontier is the exploration of ultrastable glasses (UG), amorphous states that lie much deeper in the energy landscape than those obtained by conventional cooling~\cite{ediger2017, rodriguez2022}. Studying UG is fundamental from both theoretical and technological perspectives, as it sheds light on the nature of the glass transition and amorphous states while enabling the design of mechanically robust glassy materials for engineering applications~\cite{ediger2017}. Understanding the computational routes by which such states can be generated and manipulated is the guiding question of this review.

\begin{figure*}[!th]
    \centering
        \includegraphics[width=\textwidth]{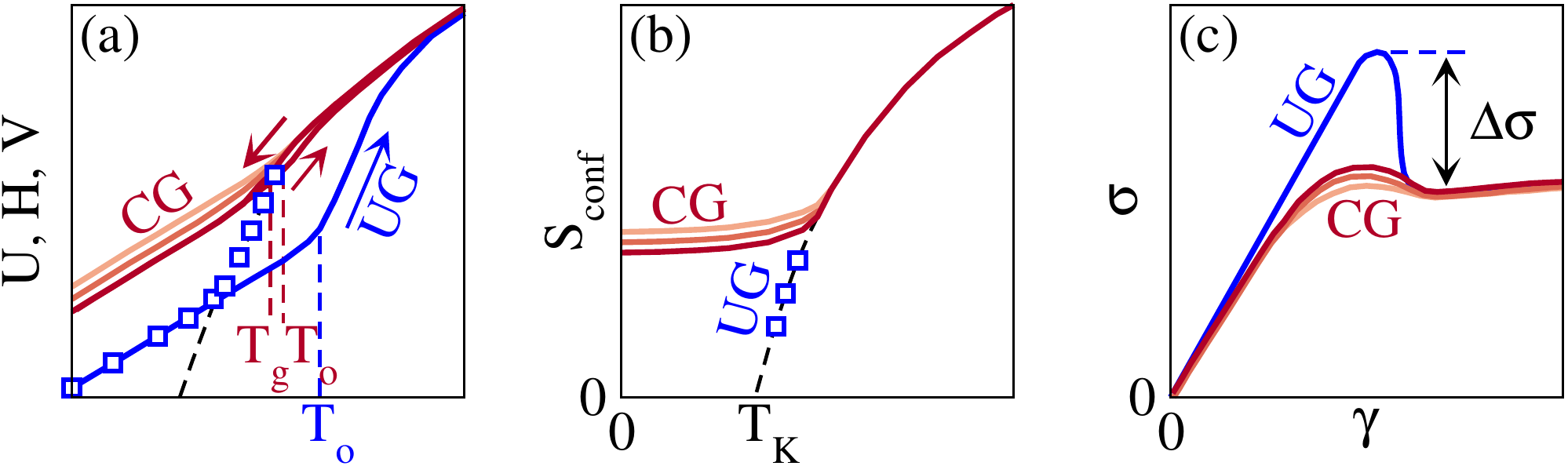}
    \caption{(a) Kinetic ($T_o$) signatures of stability as represented by internal energy $U$, enthalpy $H$, or volume $V$ versus temperature during heating ramps (right-up arrows), compared with cooling for a conventional glass (left-down arrow). $T_g$ and $T_o$ denote the glass transition temperature and the onset temperature, respectively.
    (b) Thermodynamic signature of stability as represented by the configurational entropy $S_{\rm conf}$ versus temperature $T$ for a conventional glass (CG, dark red for low cooling rates and light red for high cooling rates) and an ultrastable glass (UG, blue squares). $T_{\rm K}$ denotes the Kauzmann temperature. 
    (c) Mechanical signature of stability as represented by the stress--strain curve ($\sigma$ versus $\gamma$). The stress overshoot $\Delta\sigma$ is shown for a conventional glass (CG) and an ultrastable glass (UG).
    }
    \label{fig:fig1}
\end{figure*}

A necessary starting point is to clarify what is meant by \emph{stability} in a system that is intrinsically out of equilibrium. 
Unlike thermodynamically equilibrium materials, glasses require an operational notion that captures how long they persist, how deep they lie in the landscape, and how they respond to perturbations.
Accordingly, stability is commonly characterized through three complementary perspectives, i.e., kinetic, thermodynamic, and mechanical, summarized schematically in Fig.~\ref{fig:fig1}.

The glass transition observed during the fast cooling of a liquid, is phenomenologically marked by a dramatic increase in viscosity, reaching values of order $\eta \approx 10^{12}\,\mathrm{Pa\,s}$ beyond which the material behaves as a rigid disordered solid on experimental timescales ~\cite{angell1995}. This kinetic arrest implies that the stability of conventional glasses is inherently limited by the time available for equilibration during preparation: as the glass transition is approached, relaxation times grow so rapidly that the system becomes trapped in out-of-equilibrium configurations before it can fully explore its energy landscape \cite{angell2000}.

The kinetic stability of a glass is naturally assessed through its response to heating protocols.~\cite{swallen2007,kearns2010}
A more stable glass resides deeper in the energy landscape and thus requires more time and thermal energy to relax. In practice, kinetic stability is quantified by the onset temperature $T_o$ at which, during heating, physical quantities such as energy, enthalpy, or volume recover their equilibrium values, with higher $T_o$ indicating greater stability, as shown in Fig.~\ref{fig:fig1}a. $T_o$ also slightly increases with increasing heating rate. Originally introduced for conventionally prepared glasses, where slower cooling generically yields more stable states, this characterization remains equally valid for ultrastable glasses obtained through any preparation route.

From a thermodynamic perspective, stability reflects how deep a configuration lies in the potential or free-energy landscape \cite{debenedetti2001,sciortino2005,gupta2019}. 
More stable glasses correspond to states trapped in deeper minima, characterized by lower energies and a reduced number of accessible configurations. 
In this picture, supported by theoretical frameworks that attribute a thermodynamic origin to glass formation \cite{kirkpatrick1987, mezard1999, lubchenko2007, biroli2012}, the configurational entropy $S_{\rm conf}$ decreases as the system explores progressively deeper regions of the landscape, as illustrated in Fig.~\ref{fig:fig1}(b). 
Extrapolating this trend suggests the Kauzmann temperature $T_{\rm K}$ \cite{kauzmann1948}, where the configurational entropy would supposedly vanish, defining the hypothetical ideal glass.
Thus, the lowest value of $S_{\rm conf}$ achieved can serve as a measure of thermodynamic stability.

Finally, stability can be probed through mechanical response. 
Under applied strain, glasses initially deform elastically before yielding through irreversible rearrangements. 
More stable glasses typically exhibit larger elastic moduli and higher stress overshoots $\Delta\sigma$, reflecting an increased resistance to plastic flow \cite{berthier2025}. 
This behavior is illustrated in Fig.~\ref{fig:fig1}(c), where the ultrastable glass shows a pronounced peak of $\sigma$ compared to a conventional glass.

Taken together, these kinetic, thermodynamic, and mechanical viewpoints provide a unified picture: increasing stability corresponds to preparing amorphous states that persist longer before transforming, reside deeper in the energy landscape, and resist deformation more strongly. 

From an experimental perspective, a major breakthrough that overcame the limitations imposed by extremely long relaxation times in conventional preparation methods came in 2007, when Swallen \emph{et al.} demonstrated that physical vapor deposition (PVD) can produce glasses whose stability corresponds to thousands or even millions of years of conventional aging~\cite{swallen2007}. 
In PVD, molecules are deposited onto a substrate held near an optimal temperature, where enhanced surface mobility allows them to equilibrate efficiently before being buried by subsequent layers. 
This layer-by-layer growth enables the formation of exceptionally well-packed amorphous states.
PVD glasses became the first example of ultrastable glasses, i.e., states that display markedly lower energies, higher onset temperatures, reduced configurational entropy, and increased mechanical rigidity compared to conventionally cooled glasses.
They also exhibit improved resistance to devitrification and reduced gas permeability, properties that are technologically relevant, for example in organic electronic devices such as OLEDs~\cite{ediger2017,rodriguez2022}.

At the same time, PVD glasses strongly motivated the development of computational strategies capable of producing ultrastable glasses \emph{in silico} so allowing a microscopic understanding of the mechanisms underlying their formation. Indeed, molecular simulations provide atomistic access to the structure and dynamics of glass-forming systems, but they operate on length and time scales that are vastly shorter than those of laboratory experiments. 

Even state-of-the-art molecular dynamics simulations reach equilibrium relaxation times of at most $\tau_\alpha \sim 10^{-6}$--$10^{-3}$\,s~\cite{berthier2011,berthier2023}. Using the Maxwell relation $\eta \simeq G_\infty \tau_\alpha$, with a typical high-frequency shear modulus $G_\infty \sim 1$--$10\,\mathrm{GPa}$~\cite{rodriguez2022}, these timescales translate into effective viscosities in the range $\eta \sim 10^{3}$--$10^{7}\,\mathrm{Pa\,s}$, between those of peanut butter and tar pitch at room temperature~\cite{koop2011}. This falls dramatically short of experimental glasses, whose relaxation times near $T_g$ are of order $10^2$\,s, corresponding to the aforementioned viscosities of $\sim 10^{12}\,\mathrm{Pa\,s}$. The resulting mismatch between simulated and experimental timescales spans five to eight orders of magnitude, implying that conventional simulations probe dynamical regimes far less viscous, and hence far less stable, than those accessible in the laboratory.
A further constraint is that simulations must carefully avoid crystallization, which is particularly delicate in simple or weakly frustrated models, precisely those most commonly employed for their analytical tractability and conceptual transparency.

Overcoming these intrinsic limitations requires a change of perspective. Unlike experiments, simulations are not constrained to follow physical dynamics and can instead employ non-physical sampling strategies that accelerate exploration of configuration space while preserving statistical consistency. Over the past two decades, this idea has led to a broad family of approaches.
Early numerical efforts to reconcile experiments with simulations were strongly influenced by methods developed in the spin-glass community~\cite{swendsen1986}, leading to the adoption of parallel tempering~\cite{yamamoto2000} and mean-field-inspired approaches such as random pinning~\cite{cammarota2012,hocky2014,ozawa2015} and, more recently, random bonding~\cite{ozawa2023}. More than a decade ago, event-chain algorithms~\cite{bernard2011} pushed the boundaries of advanced Monte Carlo algorithms. In parallel, experimental advances in vapor deposition inspired computational analogues~\cite{singh2013} as well as nonequilibrium protocols such as cyclic shear~\cite{fiocco2013,parmar2019}.
A turning point came in 2017 with the optimization of swap Monte-Carlo for polydisperse systems~\cite{ninarello2017}, enabling equilibration of supercooled liquids beyond experimentally accessible regimes and yielding glasses of unprecedented stability. This development underscored the importance of enhanced phase-space exploration, effectively introducing polydispersity as an additional control dimension. 
Subsequent developments along this line have introduced grand-canonical dynamics~\cite{brito2018}, in which particle attributes such as size are allowed to evolve dynamically to facilitate equilibration, and further algorithmic refinements~\cite{ghimenti2024}.
A complementary route, based on trajectory sampling~\cite{turci2017}, targets rare dynamical fluctuations in configuration space, thereby granting access to exceptionally deep glassy states.  Finally, more recently, machine-learning-based strategies, ranging from adaptive Monte-Carlo schemes to generative models, have been explored to evaluate their potential for generating ultrastable configurations~\cite{galliano2024}. 

Some of these techniques now generate amorphous states whose stability rivals or even exceeds that of conventionally prepared experimental glasses, and in some cases approaches that of ultrastable PVD glasses. 
For this reason, they are often described as producing ultrastable glasses \emph{in silico}. 
Beyond their algorithmic significance, these methods provide a powerful route to probe deeply metastable regions of the landscape, enabling direct tests of theoretical ideas on configurational entropy, relaxation mechanisms, and mechanical response.
This review focuses on these computational strategies, their physical principles, and the extent to which they allow simulations to close the stability gap with experiments. 
Each technique is presented according to a unified framework: we first provide a concise description of the method and its underlying principles (description), followed by a detailed account of its implementation (algorithm). We typically discuss its advantages, limitations, and performance relative to alternative approaches (considerations), and conclude with an overview of its applications, highlighting its adoption and evolution within the community (applications).

We organize the review into five broad categories. The first covers physical and quasi-physical preparation protocols, such as vapor deposition~\ref{sec:PVD} and cyclic shear~\ref{sec:shear}, which are closest in spirit to experimental routes. The second category comprises equilibrium-sampling accelerators, including swap Monte Carlo (Sec.~\ref{sec:swap}), cluster moves (Sec.~\ref{sec:cluster}), and parallel tempering (Sec.~\ref{sec:tempering}), which employ unphysical yet statistically valid moves to efficiently sample the Boltzmann distribution.
The third covers phase-space and Hamiltonian modification methods, such as random pinning~\ref{sec:random_p} and bonding~\ref{sec:random_b}, which stabilize glassy states by constraining or modifying microscopic degrees of freedom. The fourth includes target-oriented and nonequilibrium methods, ranging from structural and packing optimization~\ref{sec:transient} to trajectory-ensemble approaches~\ref{sec:trajectory}. Finally, the fifth category covers machine-learning-based methods~\ref{sec:ML}, which combine and extend ideas from all previous categories. However, this categorization should not be viewed as rigid, as several algorithms admit variants that may fall into different classes, for instance, equilibrium and nonequilibrium formulations, or schemes that enforce only global balance. A dedicated section (Sec.~\ref{sec:stability}) provides a systematic comparison of the stability achieved by the different methods, summarizing results from the literature in a unified framework, complementing the primary metrics discussed above (and illustrated in Fig.~\ref{fig:fig1}).
We conclude by discussing conceptual connections between algorithms, and outline future directions toward even more efficient routes to ultrastable glasses.


\section{Vapor deposition}
\label{sec:PVD}

\noindent{\textbf{\textit{Description}:}}
Inspired by an established experimental technique \cite{swallen2007}, physical vapor deposition (PVD) involves depositing particles from a vapor onto a cold substrate, where controlled tuning of the deposition rate and substrate temperature enables the formation of ultrastable films.

\begin{figure}[!t]
    \centering
    \includegraphics[width=0.99\linewidth]{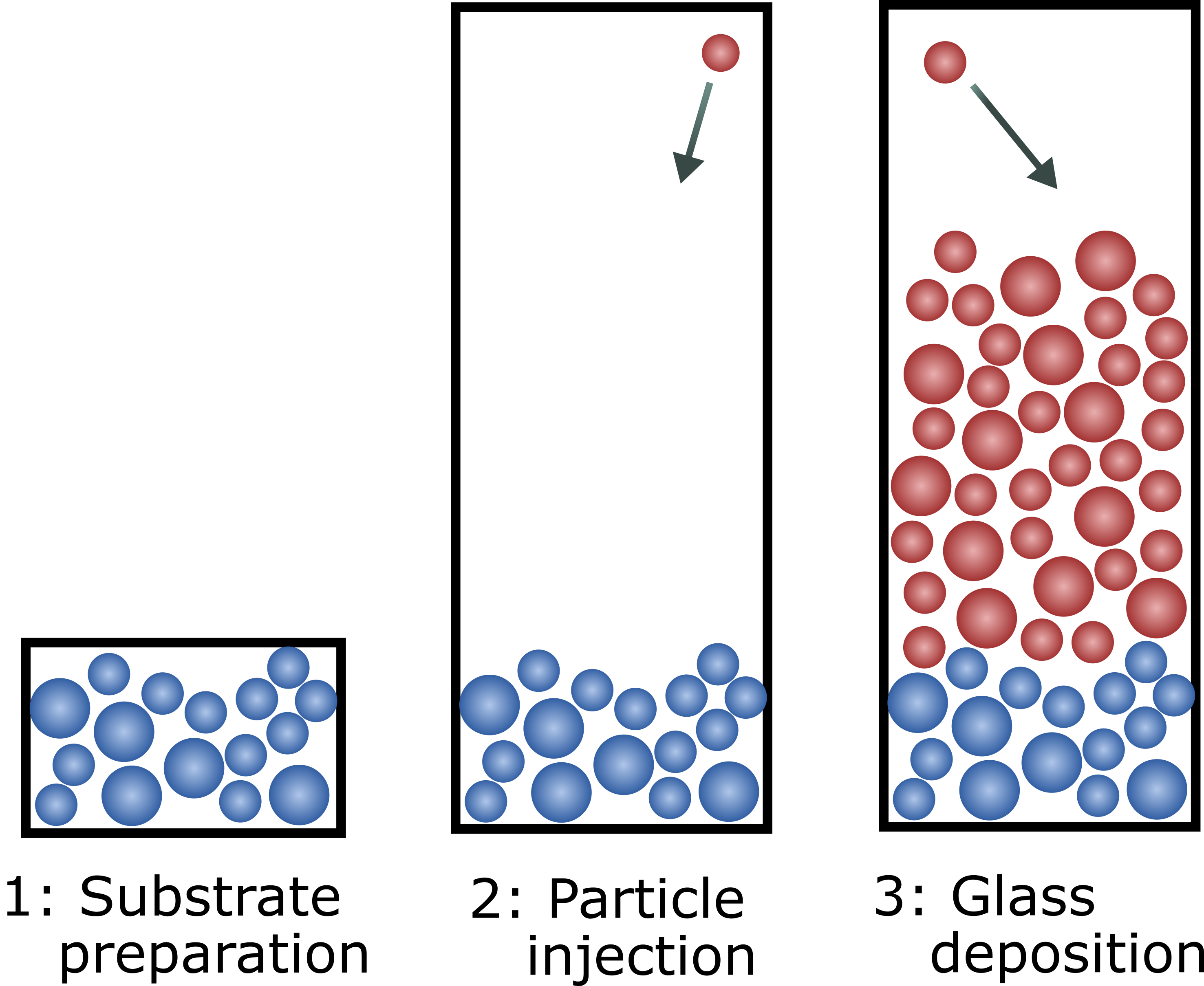}
    \caption{Schematic representation of PVD. Blue and red are substrate and deposited particles, respectively.}
    \label{fig:PVD}
\end{figure}

\vspace{0.25cm}
\noindent{\textbf{\textit{Algorithm}:}}
Begin by preparing the substrate: {\it (1)} take a bulk configuration with box size larger than spatial correlations (typically larger than $\sim 10\sigma$, with $\sigma$ the particle's size). Therefore, quench the bulk configuration below its glass-transition temperature ($T_g$) in NPT ensemble at zero pressure in preparation of its exposure to vacuum, {\it (2)} extract an $xy-$slab whose thickness in the $z$ (deposition) direction exceeds the potential cutoff (see panel 1 of Fig.~\ref{fig:PVD}), and {\it (3)} embed this slab in an elongated simulation box exposed to vacuum, with open boundaries along $z$ and periodic boundaries along $x$ and $y$ (see panel 2 of Fig.~\ref{fig:PVD}). {\it (4)} Next, initialize the deposition loop: inject with rate $\gamma$ one or a small group of particles at a time from the top with velocities sampled from a high-temperature Maxwell distribution (considering also rotational degrees of freedom for non spherical molecules) (see panel 3 of Fig.~\ref{fig:PVD}). The deposition rate $\gamma$, defined as the thickness of the deposited layer divided by the elapsed time, is an explicit control parameter of the algorithm together with the temperature of the substrate. Newly inserted particles are propagated in the NVE ensemble, while the pre-existing substrate is thermostatted in NVT in the range $0.8T_g-0.9 T_g$. Deposited particles are usually evolved in the NVE ensemble, since they thermalize to the substrate temperature between arrivals. Continue iterating insertion–integration cycles until the target film thickness is reached.

\vspace{0.25cm}
\noindent{\textbf{\textit{Considerations}:}}
The mechanism behind vapor deposition lies in the enhanced surface mobility of glass-forming materials \cite{stevenson2008,zhu2011,daley2012,brian2013,lyubimov2013,zhang2016,reid2016,zhang2017b,berthier2017b,samanta2019,moore2019,harrowell2019,leoni2023crystal,leoni2023}. At the free surface, particles experience drastically faster diffusion than in the bulk remaining mobile for several bulk relaxation times \cite{swallen2007,sun2017}, allowing them to explore configuration space and relax into low-energy states before being buried by subsequent depositions. The dynamic acceleration extends a few molecular layers into the film, its penetration length being significantly larger than that of the structural inhomogeneities induced by the interface, so that the resulting ultrastable glass is formed in the subsurface region \cite{leoni2023}. 
Consistently, the degree of ultrastability correlates with the ratio between surface, $\tau_{\rm surf}$, and bulk, $\tau_{\alpha}$, relaxation times \cite{sepulveda2014}. Indeed, the most stable glass that can be prepared at a given substrate temperature is thought to be the equilibrium supercooled liquid and its kinetic stability is characterized by $\tau_{\alpha}$. Accordingly, the maximum kinetic stability achievable for a specific material is constrained by the value of $\tau_{\rm surf}/\tau_{\alpha}$, with lower ratios favoring enhanced stability. This ratio has been found to positively correlate with liquid fragility in several studies \cite{sepulveda2014,rodriguez2022,leoni2024} suggesting that strong liquids are less effective at forming ultrastable glasses. Despite this trend, theoretical calculations~\cite{stevenson2008} indicate that liquids spanning the full range of fragilities are, in principle, capable of producing highly stable glasses. As a result, fragility by itself does not constitute a universal predictor of glass stability \cite{sepulveda2014}. Other favorable factors include low deposition rates \cite{swallen2007} (allowing equilibration at the surface), optimal substrate temperatures near $0.85\,T_g$ \cite{swallen2007}, and molecular shape influencing the anisotropy \cite{lin2004,lin2007,yokoyama2008,yokoyama2011,kim2012,dalal2015,jiang2016,antony2017,walters2017}, 
or orientational ordering, which can further enhance packing efficiency and kinetic stability. More recently, the elasticity of the substrate has been experimentally investigated as an additional parameter to enhance stability \cite{luo2024}.

A first important aspect of the substrate is its structure, which can influence the formation of the deposited layer depending on the material used. In experiments employing organic molecules such as indomethacin, the substrate is typically a crystalline solid, most commonly silicon wafers, which provide smooth and thermally stable surfaces for controlled deposition \cite{ediger2017}. Similarly to experiments, organic molecules can be deposited on a crystal surface in simulations to form an ultrastable glass \cite{sokolov2026}.
However, when considering atoms, symmetrical or simple molecules, crystalline substrates can favor heterogeneous nucleation over a wide range of conditions \cite{lupi2014}. In contrast, disordered substrates restrict nucleation to a narrower temperature range above $T_g$, and tend to remain homogeneous, at least for simple liquids \cite{leoni2023crystal}.

The typical control parameters optimized in simulations are the substrate temperature and the deposition rate.
The optimal substrate temperature is usually located between $0.8-0.9\,T_g$, similar to what is found experimentally.
The deposition rates attainable in simulations (of the order of $\gamma\sim 10^7$ nm/s) are order of magnitude larger than experiments ($\gamma\sim 0.1-1$ nm/s for organic glasses \cite{rodriguez2022}), and this is the major reason why simulations can only attain moderate ultrastability: while the gain in stability for experimental vapor deposited glasses over annealed glasses is of $\sim 8$ orders of magnitude, for simulations it reduces to $\sim 3$ orders of magnitude \cite{zhang2017,berthier2017}.


Finite system sizes further restrict the bulk region of the film which is not affected by surface effects, and limit the exploration of film-thickness-dependent properties such as those observed in experiments for deposited layers as thick as $60-70$nm \cite{jin2021}, while typical simulated layers reach a thickness of $\sim 0.1$nm. 

At temperatures close to $T_g$, diffusion from the substrate may occur; in such cases, harmonic springs should be applied to confine the substrate molecules.

The properties of the deposited species (atoms or molecules) play a key role in determining the characteristics of the resulting layer, including its structural complexity. For spherically symmetric interactions, the deposited structure is, on average, isotropic \cite{leoni2023}. In contrast, asymmetric molecules, such as those forming organic glasses like TPD, tend to produce anisotropic molecular packing \cite{ediger2017,rodriguez2022}. Another structural feature influenced by the nature of the deposited species is the development of porosity: porous structures have been observed, for instance, in films formed by silica molecules \cite{shcheblanov2024} and by water \cite{amato2025}.

Despite the limited stability range achievable in simulations, vapor deposition is the main algorithmic route to glass formation with a direct experimental counterpart, providing a unique bridge between computational and laboratory studies of ultrastable glasses.

\vspace{0.25cm}
\noindent{\textbf{\textit{Applications}:}}
The concept of producing ultrastable glasses by physical vapor deposition (PVD) originated from experiments by Ediger and co-workers, who demonstrated that vapor-deposited organic films could reach equilibrium states equivalent to those of glasses aged for millennia \cite{lyubimov2013}. This finding immediately raised fundamental questions about how far out-of-equilibrium systems could approach equilibrium by tuning kinetic pathways during formation.

Shortly thereafter, computer simulations began to reproduce and rationalize these observations. De Pablo and collaborators implemented atomistic and coarse-grained deposition protocols, showing that controlled deposition on cold substrates indeed leads to glasses of enhanced stability \cite{singh2011,singh2013,lyubimov2013,lin2014,lyubimov2015,dalal2015,reid2016,walters2017,seoane2018b}. Since then many works explored the connections between vapor deposition and theoretical ideas of glass equilibration via surface mobility, and tried to extend the range of systems capable of forming ultrastable glasses, including low-fragility liquids \cite{sepulveda2014} and tetrahedral, network-forming, materials \cite{leoni2024}.
Other works have explored the structural properties of deposited glasses and their connection to bulk quantities such as viscosity and relaxation dynamics. A related but much less explored route is the formation of ultrastable glasses by precipitation from solution \cite{douglass2019}, where a glass-forming solute grows at a solvent interface and the enhanced interfacial mobility enables the formation of highly stable amorphous structures.

\section{Cyclic shear}
\label{sec:shear}

\noindent{\textbf{\textit{Description}:}}
Under cyclic, or oscillatory, shear deformation, the energy of a glass can either decrease or increase depending on key parameters such as strain amplitude \cite{fiocco2013,leishangthem2017}. 
Similar to the compaction of granular materials under cyclic shear \cite{pouliquen2003}, tapping \cite{knight1995}, or compression–decompression cycles \cite{kumar2016}, the deformation amplitude must be carefully chosen to enhance stability by balancing overaging and rejuvenation effects \cite{lacks2004}.

\begin{figure}[!t]
    \centering
    \includegraphics[width=0.7\linewidth]{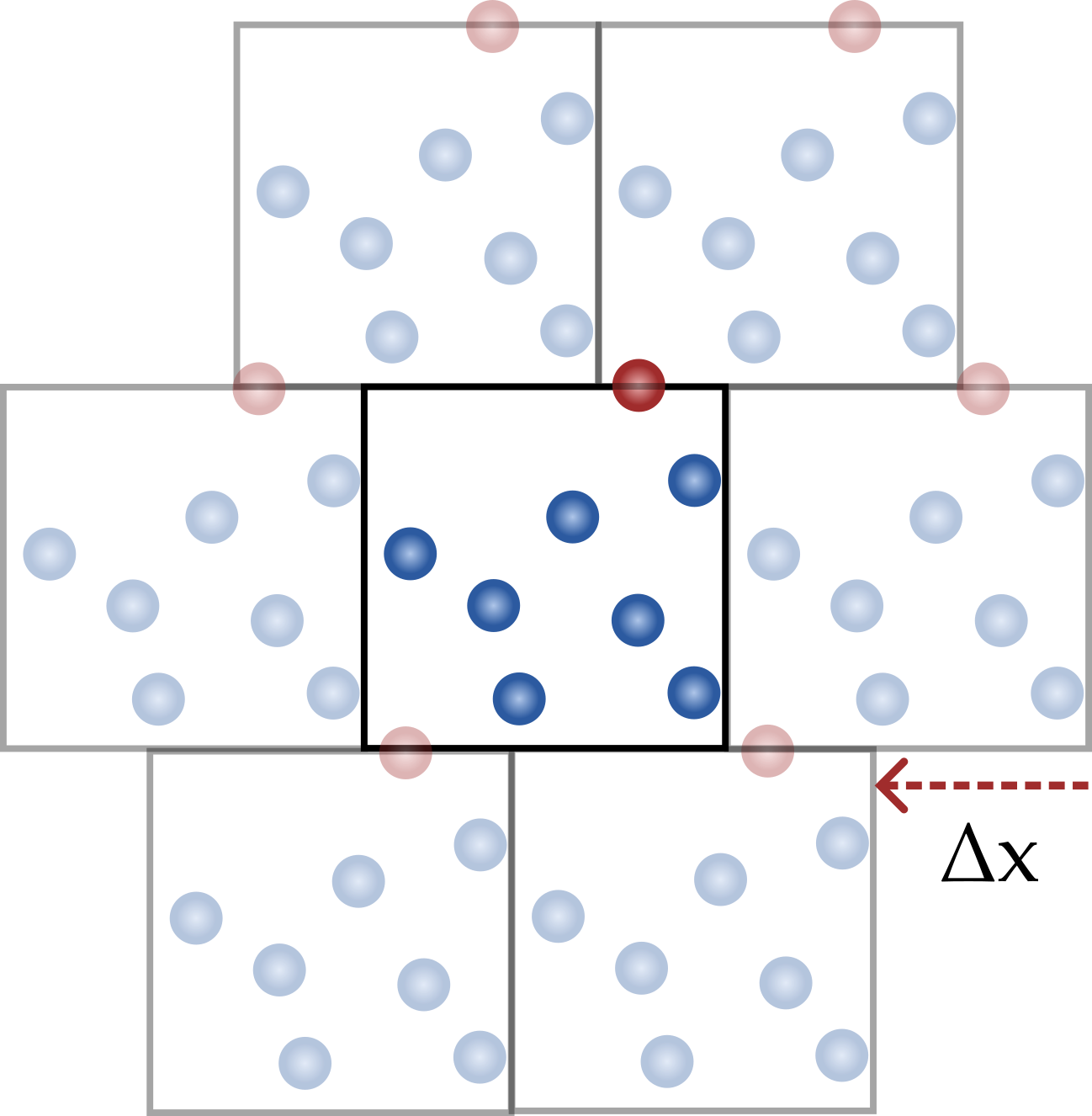}
    \caption{Schematic illustration of cyclic shear applied to a glass-forming system under Lees–Edwards periodic boundary conditions. 
    }
    \label{fig:shear}
\end{figure}

\vspace{0.25cm}
\noindent{\textbf{\textit{Algorithm}:}}
To prepare a glass with enhanced stability using cyclic shear (schematically illustrated in Fig.~\ref{fig:shear}), one typically proceeds as follows:
(1) initialize a periodic simulation box containing $N$ particles and quench the system to obtain a glass;
(2) impose shear deformation in a strain-controlled setting with appropriate boundary conditions, such as Lees--Edwards periodic boundary conditions \cite{lees1972,weik2019,bindgen2021};
(3) apply an oscillatory strain of amplitude $\gamma_{\rm max}$ slowly in a quasi-static manner.

The energy is typically monitored stroboscopically each time the strain $\gamma$ returns to zero. It decreases as the number of cycles increases when $\gamma_{\rm max}$ is below yielding value. The final configuration at $\gamma=0$ is then used to assess the stability of the glass.

\vspace{0.25cm}
\noindent{\textbf{\textit{Considerations}:}}
Deformation, such as shear, is widely used to probe the mechanical properties of materials, including steady-state flow \cite{bonn2017}, oscillatory response \cite{knowlton2014}, shear-band formation \cite{parmar2019,golkia2020}, failure of amorphous solids \cite{schuh2007,procaccia2017}, shear-melting and resolidification \cite{perez2018}, and the yielding transition \cite{leishangthem2017,kawasaki2016,regev2015,berthier2025,lin2014b,liu2016,das2020}.

When the applied shear strain is sufficiently large, the glass yields and enters a flowing state. Interestingly, however, when cyclic shear is applied with an amplitude below the yielding point, it can anneal the system and enhance its stability \cite{fiocco2013,priezjev2018,priezjev2019,leishangthem2017}. The underlying reason is that shear-induced plastic rearrangements can progressively drive the system toward lower-energy states under periodic deformation. 
As the number of cycles increases, the system approaches a steady state. The final energy level depends on the strain amplitude $\gamma_{\rm max}$: the energy decreases with increasing $\gamma_{\rm max}$ up to the yielding point, above which the system yields and undergoes rejuvenation. Therefore, the strain amplitude must be carefully tuned to maximize the annealing effect.
However, for a given shear protocol, cyclic annealing cannot reduce the energy below a certain threshold. To reach deeper energy levels, alternative annealing methods, such as thermal annealing, must be employed~\cite{bhaumik2021,yeh2020}.

\vspace{0.25cm}
\noindent{\textbf{\textit{Applications}:}}
Most studies of cyclic shear have been performed either using athermal quasi-static shear (AQS) simulations or finite-strain-rate simulations at very small shear rates~\cite{fiocco2013,leishangthem2017,bhaumik2021,yeh2020}, with only a few exceptions considering finite-temperature, finite-strain-rate protocols such as SLLOD~\cite{parmar2019}.
In AQS simulations, small strain increments are applied, each followed by energy minimization~\cite{maloney2006}, and these two steps are repeated iteratively. This protocol isolates purely mechanical responses, such as plastic events, avalanches, and yielding, from thermal effects, thereby revealing how strain drives transitions between energy minima in a complex energy landscape.

Building on these insights, cyclic or oscillatory shear was first studied in the context of memory encoding, whereby the system retains information about the training amplitude in the energy landscape~\cite{fiocco2013}. It was later exploited as a means of annealing glasses by tuning the strain amplitude~\cite{leishangthem2017,parmar2019}.
Recent work~\cite{krishnan2023} showed that multidirectional oscillatory shear reaches lower steady-state energies than unidirectional shear, suggesting that carefully designed shear cycles can anneal glasses and enhance their stability rather than degrade it.

A shear protocol applied to soft glass suspensions \cite{bantawa2025} found a connection between stability and microscopic mechanical features such as narrowing and symmetrization of local stress distributions, a trend also observed in polydisperse glassformers \cite{leoni2025}, reinforcing the link between stability and stress homogenization.

Hyperuniform states, which may be associated with enhanced glass stability \cite{yanagishima2021,fan2026}, have been observed in cyclically driven glasses below yielding, whereas above yielding the system exhibits enhanced heterogeneity and fluctuations \cite{mitra2021}.

Recent works have explored the connection between shear deformation and active forces in relation to glass stability. In Ref.~\cite{sharma2025}, internal activity in amorphous solids is shown to play a role similar to external shear in controlling glass stability. Moderate activity anneals the system into deeper energy minima, thereby enhancing stiffness, whereas excessive activity fluidizes the system and erases memory of the preparation history. This activity-induced tuning also drives a ductile-to-brittle transition, mirroring shear-driven glasses: low activity leads to homogeneous flow, whereas high activity promotes shear-band-mediated failure.
Priya et al.~\cite{priya2025} emphasized the critical role of activity in shaping the mechanical memory of ultrastable glasses, particularly through its regulation of shear-band formation and evolution. 

Cyclic shear is also relevant to experiments, particularly in soft-matter rheology~\cite{denisov2015} and fatigue failure in metallic glasses~\cite{jia2018}.


\section{Swap Monte-Carlo}
\label{sec:swap}

\noindent{\textbf{\textit{Description}:}}
The Swap Monte-Carlo (SMC) algorithm extends standard Monte-Carlo dynamics by supplementing local particle displacement moves with random diameter-swap moves, both accepted or rejected according to the Metropolis criterion to preserve detailed balance.
SMC has demonstrated high efficiency in simulations of simple glass-forming mixtures of multicomponent and polydisperse particles.\\

\begin{figure}[!t]
    \centering
    \includegraphics[width=0.99\linewidth]{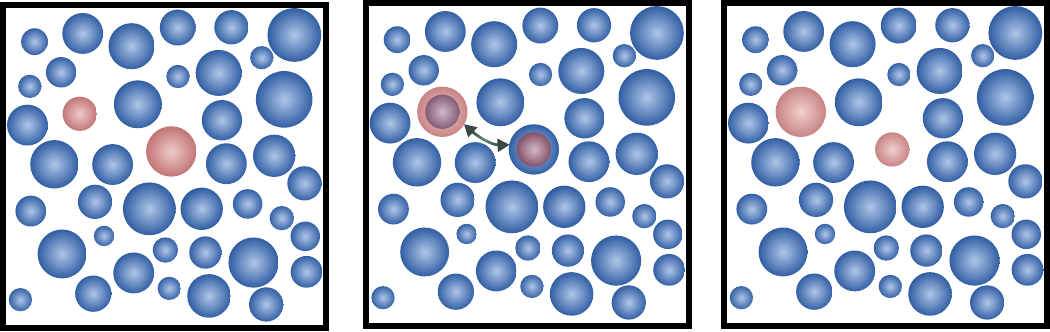}
    \caption{Schematic representation of the Swap Monte-Carlo (SMC) algorithm. In addition to standard particle displacement moves, pairs of particles are randomly selected and their diameters are exchanged with a given probability. The proposed swap is accepted or rejected according to the Metropolis criterion based on the resulting energy change. }
    \label{fig:swap}
\end{figure}

\noindent{\textbf{\textit{Algorithm}:}}
SMC enables the generation of equilibrium simulations of supercooled liquids, which can subsequently be driven out of equilibrium to form glasses through rapid cooling or compression. 
During SMC simulations, at each Monte-Carlo step, in addition to standard particle displacement moves, a swap move is attempted with probability $p$. A swap move proceeds as follows:
(1) Two particles are randomly selected, with diameters $\sigma_i$ and $\sigma_j$. (2) The particle diameters are exchanged and the resulting energy difference between the new and old configurations is computed, $\Delta U = U_{\mathrm{new}} - U_{\mathrm{old}}$. (3) The move is accepted or rejected according to the Metropolis criterion,
\begin{equation}
p_{\rm acc}(i \leftrightarrow j) = \min\{1, \exp(-\beta \Delta U)\}.
\end{equation}\\
Those three steps are illustrated schematically in Fig~\ref{fig:swap}.

\noindent{\textbf{\textit{Considerations}:}}
To ensure that crystallization does not occur, throughout the simulation, simple observables such as the potential energy must be monitored. Structural relaxation times $\tau_\alpha$ can then be extracted from time-dependent correlation functions, such as the self-intermediate scattering or overlap function, as a function of temperature or density. These relaxation times are used both to verify equilibration and to quantify the dynamical speedup achieved relative to standard MC simulations.

An equivalent implementation of the SMC algorithm can be obtained by exchanging particle positions rather than diameters. However, this approach prevents the tracking of single-particle dynamics, since particles undergo arbitrarily large effective displacements during swap moves.

The SMC algorithm constitutes one of the most significant advances in the numerical study of glass-forming systems and supercooled liquids. 
Using Metropolis-accepted displacement and swap moves, SMC preserves detailed balance while enabling non-local exploration of configuration space.
This simple yet effective approach establishes SMC as the most efficient known method for equilibrating polydisperse mixtures across a broad range of dimensions, owing to its independence from geometric constraints. It yields an enormous dynamical speedup, with the most effective systems exhibiting estimated gains of at least ten orders of magnitude, though accurate quantification remains challenging, as it requires extrapolating standard dynamical timescales into thermally inaccessible regimes, and the true speedups could in fact be considerably larger.

The algorithm’s efficiency can be improved by adjusting the swap fraction $p$ to reduce relaxation times and limiting swaps to particles with diameter differences smaller than an optimized value.
Using highly polydisperse or carefully designed discrete (e.g., ternary) particle size distributions is essential to suppress crystallization and achieve a large dynamical gain. For continuous polydisperse systems, the optimal polydispersity for hard spheres and repulsive particles is approximately 23\% of polydispersity~\cite{ninarello2017}, while discrete mixtures can achieve comparable performance through optimized composition and size ratios~\cite{gutierrez2015,parmar2020,jung2023}.
Moreover, as a Monte-Carlo–based method, its stochastic and inherently serial nature makes large-scale parallelization nontrivial~\cite{berthier2019c}.

\noindent{\textbf{\textit{Applications}:}}
The method has its origins in early lattice MC studies of binary alloys, where swap-like exchanges were used to compute order parameters and phase transitions.~\cite{fosdick1959} 
It was later generalized to off-lattice systems such as Lennard-Jones microclusters~\cite{tsai1978} and hard-sphere mixtures.~\cite{gazzillo1989} 
The first applications to glass-forming liquids employed binary mixtures, demonstrating the potential of swap dynamics to reach deeply supercooled states.~\cite{grigera2001}
Its algorithmic efficiency has also made it a valuable tool for mapping phase diagrams~\cite{fernandez2007, gutierrez2015} and testing theoretical frameworks such as the random first order transition theory~\cite{biroli2008}. However, these early binary models were later shown to crystallize easily~\cite{brumer2004}.
A 2017 reassessment of SMC~\cite{ninarello2017} identified optimal parameters (particle size distribution, softness, and nonadditivity) that suppress crystallization and yield extremely high equilibration speedups, redefining glass simulation standards.

One of the key achievements of SMC is its contribution to understanding the entropy crisis in glass formation, as it enabled the observation of a pronounced decrease in configurational entropy over an unprecedented cooling range~\cite{berthier2017} and, in two dimensions, hinted at a possible zero-temperature transition characterized by vanishing entropy and a growing static length scale~\cite{berthier2019}.
SMC also provides insights into the evolution of the rugged energy landscape upon deep supercooling~\cite{nishikawa2022}.

SMC has been instrumental in elucidating the vibrational properties of ultrastable glasses, revealing that quasi-localized modes (soft excitations distinct from phonons) govern low-frequency vibrations, thermal anomalies, and tunneling phenomena~\cite{wang2019, khomenko2020, khomenko2021, wang2022}. These soft modes generally follow $D(\omega) \sim \omega^4$, with deviations linked to stability and system size~\cite{wang2022, schirmacher2024}. Sound attenuation serves as a key probe of these excitations, with more stable glasses exhibiting reduced damping and delayed quartic scaling ~\cite{wang2022, szamel2022}. Furthermore, tunneling two-level systems, long associated with glass anomalies, are increasingly linked to soft quasi-localized modes, though their connection remains subtle and nonlinear~\cite{scalliet2019, khomenko2020, khomenko2021}.

SMC has been as well crucial for exploring thermodynamic transitions predicted by high-dimensional mean-field glass theory.
In three-dimensional hard-sphere glasses, simulations have revealed signatures of a Gardner-like transition~\cite{seoane2018,jin2018}, at which the glass free-energy basin fragment into a hierarchically organized, marginally stable landscape. However, softer particle models show no clear evidence~\cite{scalliet2017,scalliet2019}, and in two dimensions the phenomenon appears only as a strong crossover~\cite{liao2019}.
Moreover, simulations across dimensions $d=3$–$10$ reveal finite-dimensional remnants of mean-field Ising-like spinodal criticality, strongly softened in low dimensions and gradually approaching mean-field behavior as $d$ increases~\cite{berthier2020}.

The rheological behavior of glasses, including their yielding and failure, has been extensively studied through athermal quasi-static simulations (see Sec.~\ref{sec:shear} for further details), with SMC enabling access to ultrastable configurations. These simulations revealed a brittle, discontinuous yielding transition with macroscopic failure reminiscent of experimental observations~\cite{ozawa2018}. Further work suggested that the crossover from brittle to ductile behavior may correspond to a critical point~\cite{ozawa2020}, though this interpretation remains debated~\cite{richard2021}. Recent studies have also examined cyclic deformation~\cite{yeh2020}, and plastic events~\cite{ozawa2022} to deepen understanding of this transition.

The role of the SMC algorithm in the glass transition debate has generated significant theoretical interest, as it has been used as a probe to distinguish between competing theories of glass formation. One line of interpretation holds that SMC's efficiency undermines thermodynamic, cooperative theories such as RFOT, since altering purely local dynamical rules dramatically changes the relaxation time~\cite{wyart2017}. However, several works push back on this view: within RFOT, the speedup can be understood as a postponement of the onset of glassy dynamics through ''crumbling metastability,'' leaving the underlying free-energy landscape intact~\cite{berthier2019d}. At the mean-field level, replica liquid theory confirms that SMC shifts the dynamical transition point relative to standard Monte-Carlo, suggesting a modification rather than a refutation of thermodynamic glass theory~\cite{ikeda2017}. More explicitly, SMC dynamics can be governed by an effective potential that stabilizes configurations at lower energies, shifting the glass transition to lower temperatures, with the magnitude of the effect tied to polydispersity~\cite{brito2018}. A mode-coupling-inspired framework further shows that size swaps open an additional relaxation channel for density fluctuations, moving the dynamic glass transition to higher volume fractions in hard-sphere mixtures~\cite{szamel2019}. Simulations have further confirmed this shift in the mode-coupling temperature induced by SMC~\cite{kuchler2023}. A unifying perspective comes from the notion of time-reparametrization softness~\cite{ghimenti2026}: local constraints and global landscape complexity are not mutually exclusive but complementary as mean dynamics is invariant under time reparametrization, while the landscape governs the structure of correlations, and SMC-like algorithms exploit precisely this softness. Taken together, these works suggest that the sensitivity of the glass transition to dynamical rules reflects not the irrelevance of thermodynamic complexity, but a subtle interplay between local kinetics and the free-energy landscape.


\section{Event-Chain Cluster moves}
\label{sec:cluster}

\noindent{\textbf{\textit{Description}:}}
One of the central advantages of Monte-Carlo simulations over Molecular Dynamics is the possibility of employing non-physical collective or \emph{cluster moves} that dramatically accelerate relaxation. A wide variety of cluster moves have been developed for fluids and complex systems. In what follows, we focus on the family of \emph{lifted} Markov Chain algorithms~\cite{chen1999,bernard2009}, which have recently shown great promise for tackling exceptionally hard sampling problems, including glass-forming systems~\cite{ghimenti2024}. For simplicity, we limit our discussion to the case of hard interactions, but generalization to arbitrary potentials are available~\cite{michel2014,krauth2021,nishikawa2025}.

\vspace{0.25cm}
\noindent{\textbf{\textit{Algorithm}:}}
Lifted Monte-Carlo algorithms proceed through an irreversible Markov Chain in an extended configuration space.  
We illustrate the construction for two classes of collective moves: translational updates, realized through Event-Chain Monte-Carlo (ECMC)~\cite{bernard2009}, and compositional updates, realized through the Collective Swap (cSwap) algorithm~\cite{ghimenti2024}. The two schemes can also be combined in hybrid implementations.  

\begin{figure}[!t]
    \centering
    \includegraphics[width=0.99\linewidth]{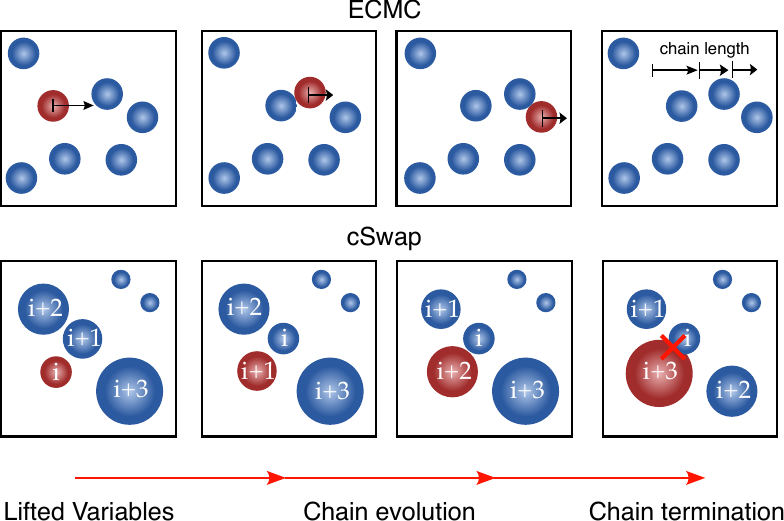}
    \caption{Schematic representation of lifted Monte-Carlo dynamics for Event-Chain Monte-Carlo (ECMC, top row) and Collective Swap (cSwap, bottom row). 
    In ECMC, the lifted variables consist of the active particle index $i$ and the propagation direction $\mathbf{d}$. 
    The active particle moves deterministically along $\mathbf{d}$ until a collision occurs, at which point the lifting rule transfers motion to the collided particle, maintaining the same direction. 
    In cSwap, the lifted variable corresponds to the index $i$ in the ordered diameter array. 
    The active particle $i$ sequentially exchanges its diameter with those of neighboring particles ($i+1, i+2, \ldots$) until an overlap event terminates the chain.
}
    \label{fig:cluster_moves}
\end{figure}

(1) Begin by defining an extended configuration space by introducing additional degrees of freedom (the \emph{lifting variables}) that govern how the dynamics unfolds (first column of Fig.~\ref{fig:cluster_moves}).  
In ECMC, these variables consist of the index $i$ of the active particle and a propagation direction $\mathbf{d}$ (the colored particle and the arrow in the top panel in Fig.~\ref{fig:cluster_moves}), while in cSwap the lifted variable is an index $i$ identifying a particle within an array where all diameters are sorted in increasing size (the colored particle in the bottom panel of Fig.~\ref{fig:cluster_moves}). 

(2) Evolve deterministically the variables according to the chosen dynamics (second column of Fig.~\ref{fig:cluster_moves}).  
In ECMC, the active particle $i$ moves along $\mathbf{d}$ with constant velocity until an event, such as a collision, occurs.  
When periodic boundary conditions are employed, the propagation direction is typically chosen from the set $\{+\mathbf{e}_x, +\mathbf{e}_y, +\mathbf{e}_z\}$, cycling through the Cartesian axes to ensure isotropic sampling.  
In cSwap, the evolution takes place in composition space rather than real space: the active particle $i$ successively exchanges its diameter with its right-hand neighbor in the sorted array, and each accepted swap immediately updates the configuration before the next attempt.

(3) Define a lifting rule specifying how activity is transferred when deterministic propagation is interrupted (third column of Fig.~\ref{fig:cluster_moves}).  
In ECMC, upon a collision, the moving particle stops and the collision partner becomes active, inheriting the same propagation direction.  
In cSwap, when a swap attempt leads to an overlap, the chain terminates and the activity label shifts to a neighboring particle, either to the left or to the right in the sorted array.

(4) Terminate each chain after a prescribed cumulative displacement (in ECMC) or after the first rejected exchange (in cSwap), forming a collective update that satisfies global balance with respect to the Boltzmann distribution (last column of Fig.~\ref{fig:cluster_moves}).  

(5) Ensure ergodicity by periodically resampling the lifting variables.  
In ECMC, select a new active particle and propagation direction after each chain, while in cSwap uniformly reassign the activity index $i$ along the ordered array with a small probability $p_r \sim 1/N$.  
Together, these steps define a complete lifted Monte-Carlo cycle capable of generating irreversible yet statistically exact trajectories through configuration space.

\vspace{0.25cm}
\noindent{\textbf{\textit{Considerations}:}}
The dynamics of the lifted variables intentionally violate detailed balance, yet obey the more general condition of global balance, ensuring that the steady-state distribution remains Boltzmann~\cite{bernard2009,kapfer2013,maggs2022,krauth2021,monemvassitis2023}. 
Although the resulting trajectories are irreversible and therefore unphysical, they often yield substantially faster equilibration than standard, reversible Monte-Carlo schemes. 

A second major advantage of the lifted approaches described above is that they are \emph{rejection-free}: all proposed micro-moves are accepted by construction, so that the acceptance rate does not decay as the system becomes dense or strongly correlated. 
In Event-Chain Monte-Carlo (ECMC), this feature leads to dramatic gains in sampling efficiency. 
For instance, ECMC achieves speedup factors on the order of $10^3$ with respect to standard Metropolis Monte-Carlo for hard-sphere systems, and remains one of the most efficient algorithms available for dense fluids~\cite{li2022}. 
In two-dimensional polydisperse disks, the ratio of relaxation times between ECMC and standard Monte-Carlo decreases from roughly $22$ to $10$ as the packing fraction increases, demonstrating that the advantage persists even near the glassy regime~\cite{ghimenti2024}.  

A similar acceleration is observed in the compositional space explored by the Collective Swap (cSwap) algorithm. 
Compared to standard swap Monte-Carlo, cSwap yields a dramatic reduction in equilibration times, with speedup factors that \emph{increase} with volume fraction, reaching values of about $40$ at the highest packing fractions studied~\cite{ghimenti2024}.
This inverted trend, opposite to that of ECMC, suggests that lifted dynamics in composition space become particularly efficient as steric constraints intensify.

Several extensions have further broadened the applicability of lifted algorithms. 
The introduction of the \emph{factorized Metropolis filter} allows ECMC to be generalized beyond hard interactions, making it applicable to continuous~\cite{michel2014}, long-range potentials~\cite{kapfer2016,faulkner2018}, and to parallel programming~\cite{li2021}.

\vspace{0.25cm}
\noindent{\textbf{\textit{Applications}:}}
The development of non-local Monte-Carlo methods traces back to the pioneering work of Swendsen and Wang~\cite{swendsen1987}, who introduced a cluster algorithm capable of suppressing critical slowing down in spin systems near continuous phase transitions. 
This demonstrated that non-physical, collective updates, if properly constructed to respect detailed balance, could drastically accelerate equilibration in strongly correlated regimes~\cite{dress1995,buhot1998,santen2000,liu2004}.

The concept of irreversibility in Monte-Carlo dynamics emerged later with the introduction of algorithms such as 
Event-Chain Monte-Carlo (ECMC)~\cite{bernard2009}: they replaced stochastic trial moves with deterministic, rejection-free propagation in an extended configuration space that satisfies global rather than detailed balance. 
Originally developed for hard-sphere and hard-disk systems, ECMC contributed to solve a long-standing problem of the nature of the liquid-to-hexatic transition in two dimensions~\cite{bernard2011}. 

The application of lifted algorithms to the study of ultrastable glasses is a more recent development~\cite{ghimenti2024,nishikawa2025}. 
The collective Swap (cSwap) algorithm extends the idea of event-driven, rejection-free dynamics to composition space. 
Together, these advances mark the convergence of ideas from critical phenomena, nonequilibrium statistical mechanics, and glass physics into a unified framework of lifted Monte-Carlo algorithms for the efficient sampling of complex free-energy landscapes.


\section{Parallel tempering}
\label{sec:tempering}

\noindent{\textbf{\textit{Description}:}}
Parallel tempering is an advanced Monte-Carlo sampling method that runs multiple simulations in parallel at different temperatures, enabling the system to overcome kinetic barriers and more efficiently explore the complex energy landscape of supercooled liquids.

\begin{figure}[!t]
    \centering
    \includegraphics[width=0.99\linewidth]{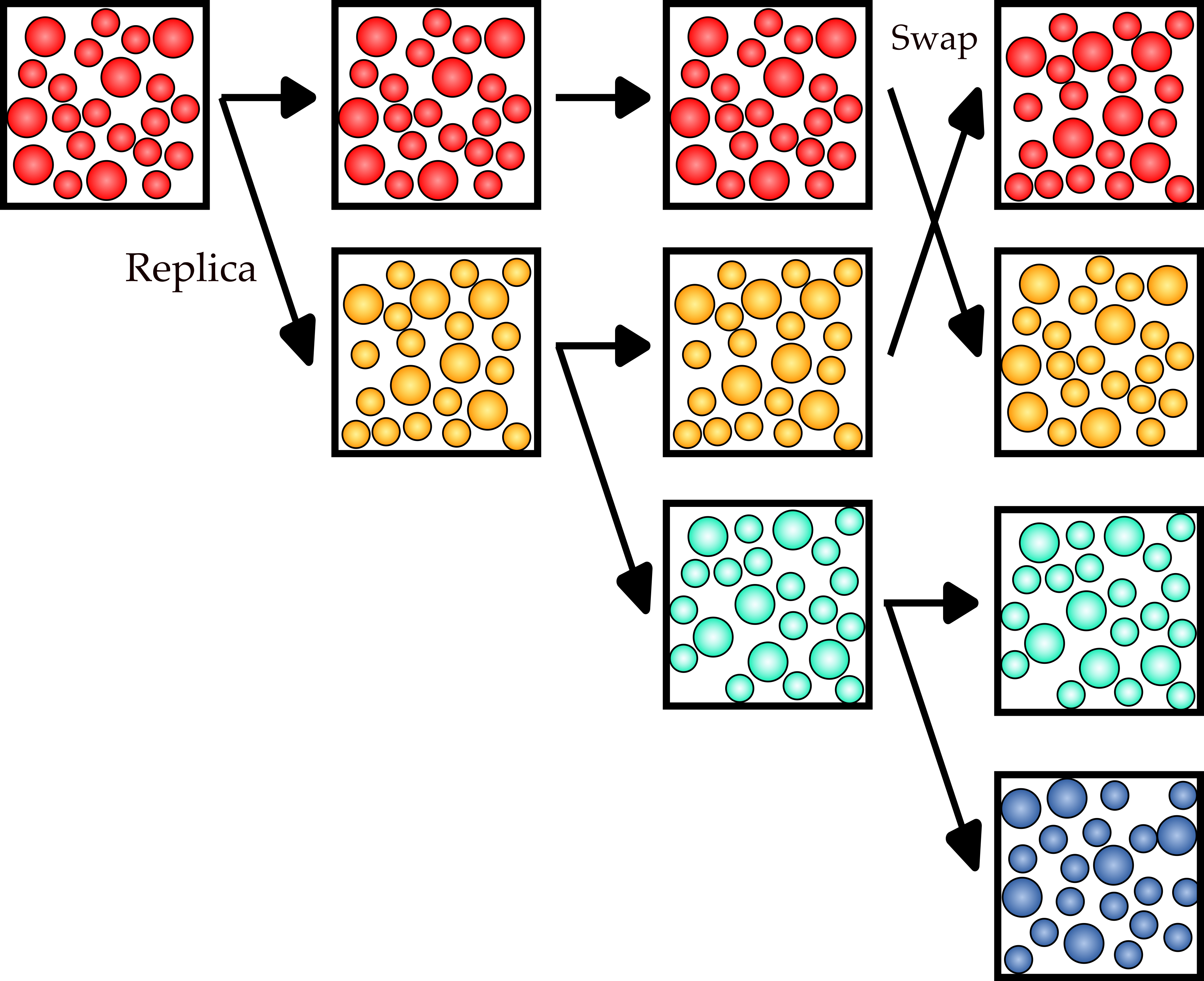}
    \caption{Schematic representation of the Parallel Tempering algorithm. From left to right: initialize the system, generate replicas at different temperatures, construct a temperature ladder, and perform configuration exchanges between replicas to enhance sampling efficiency.}    \label{fig:PT}
\end{figure}

\vspace{0.25cm}
\noindent{\textbf{\textit{Algorithm}:}}
The algorithm is schematically shown in Fig.~\ref{fig:PT}.
(1) Initialize the system: Begin by preparing a disordered initial configuration of the system. Launch a dynamical simulation, which can be either Monte-Carlo or molecular dynamics, at a chosen reference temperature $T_0$. At this stage, the system explores the phase space at a high enough temperature to avoid being trapped in local energy minima, ensuring a broad sampling of configurations.\\
(2) Create temperature replicas: Duplicate the current configuration and initiate a second simulation at a lower temperature $T_1 < T_0$. Each replica now evolves independently according to its assigned temperature.\\
(3) Build a temperature ladder: Continue this process iteratively, creating a series of $n$ replicas down to the target temperature $T_{n-1}$. This set of simulations at different temperatures forms a \emph{temperature ladder}, which is crucial for the parallel tempering method: higher-temperature replicas facilitate barrier crossing, while lower-temperature replicas provide detailed sampling of low-energy configurations.\\ 
(4) Perform configuration exchanges: At regular intervals of $N_{\mathrm{swap}}$ simulation steps, attempt to exchange configurations between replicas at adjacent temperatures. The swaps are governed by a Metropolis-like acceptance criterion,
\begin{equation}
        p_{\rm acc}(j \leftrightarrow j \pm 1)
        = \exp\!\left[-(\beta_j - \beta_{j \pm 1})\, \Delta U\right],
\end{equation}
where $\Delta U = E_{\mathrm{pot}}(x_j) - E_{\mathrm{pot}}(x_{j \pm 1})$ is the difference in potential energy between the two configurations being swapped, and $\beta_j = (k_B T_j)^{-1}$ is the inverse temperature of replica $j$. This exchange mechanism allows lower-temperature replicas to escape local minima by swapping with higher-temperature replicas, enhancing sampling efficiency.\\
(5) Iterate and equilibrate: Repeat the simulation and swap procedure for sufficiently long times so that all replicas can equilibrate and explore their respective configuration spaces thoroughly. The combination of independent evolution at different temperatures and periodic exchanges ensures that the overall ensemble represents the correct Boltzmann distribution at each temperature.\\

\noindent{\textbf{\textit{Consideration}:}}
Parallel tempering is a powerful sampling method for supercooled liquids, allowing equilibration at very low temperatures and thereby enabling the production of stable glassy states. Its key advantage is the ability to cross large energy barriers by exchanging configurations between replicas at different temperatures. In simple terms, a configuration trapped in a narrow basin at low temperature can be swapped to a higher temperature, where it can freely move across the landscape into another basin, and then return to low temperature, now trapped in a different and previously inaccessible region. This leads to improved equilibration and more accurate estimation of thermodynamic and structural properties in deeply supercooled regimes. However, parallel tempering also has limitations: it requires the simultaneous simulation of multiple replicas, increasing computational cost, and the efficiency of exchanges depends sensitively on the choice of temperature ladder, with poorly chosen spacing leading to low swap acceptance rates. To optimize parallel tempering, one can carefully tune the temperature ladder to maintain moderate acceptance rates (typically $20-50\%$), balance the number of replicas against computational resources, or employ adaptive schemes that adjust temperatures dynamically during the simulation.

Another computational method, called Population Annealing, can be viewed as a version of parallel tempering in which configuration exchanges are achieved entirely via probabilistic reweighting instead of running separate simulations at each temperature. In this method a large ensemble of configurations is initialized at a high temperature $\beta_1 = (k_B T_1)^{-1}$ and gradually annealed to lower temperatures in small steps. At each step, configurations are reweighted according to $W_i = \exp[-(\beta_j - \beta_{j+1})E_{pot}(x_i)]$ and resampled proportionally to $W_i$. This enables efficient exploration of low-temperature states without the explicit need to run computationally demanding simultaneous simulations at multiple temperatures, combining the benefits of enhanced sampling and barrier crossing.

\vspace{0.25cm}
\noindent{\textbf{\textit{Applications}:}}
Parallel tempering was first introduced in the spin-glass community~\cite{swendsen1986,marinari1998}, where it also became known as the replica Monte-Carlo method~\cite{swendsen1986} or replica exchange~\cite{sugita1999}. Its earliest applications to structural glass formers involved simulations of small binary soft-sphere systems with $N=36$ particles~\cite{coluzzi1998}, later extended to significantly larger systems of up to $N=1000$ particles~\cite{yamamoto2000}. These pioneering studies reported only a limited speedup, typically on the order of one to two decades. The method became particularly attractive with the increasing availability of parallel computing resources, as the replicas at different temperatures can be simulated largely independently and efficiently distributed across multiple processors. In the context of supercooled liquids, parallel tempering was initially employed mainly for equilibration~\cite{yamamoto2000,demichele2002}, and subsequently used in investigations of random pinning~\cite{kob2013} and in measurements of the point-to-set correlation length~\cite{yaida2016,berthier2016b}. More recently, it has served as a benchmark for assessing modern enhanced-sampling strategies, including swap Monte-Carlo and normalizing-flow-based methods~\cite{jung2024}.

\section{Random pinning}
\label{sec:random_p}

\noindent{\textbf{\textit{Description}:}}
Random pinning stabilizes a system by artificially freezing a subset of particles, rather than by lowering the temperature. More specifically, a fraction of particles is randomly selected from an equilibrium supercooled liquid configuration, and their positions are permanently pinned. The stability of the remaining mobile (unpinned) particles is then examined. As the concentration of pinned particles increases, the dynamics of the mobile particles slow down and the system becomes progressively more stable.

\begin{figure}[!t]
    \centering
    \includegraphics[width=0.99\linewidth]{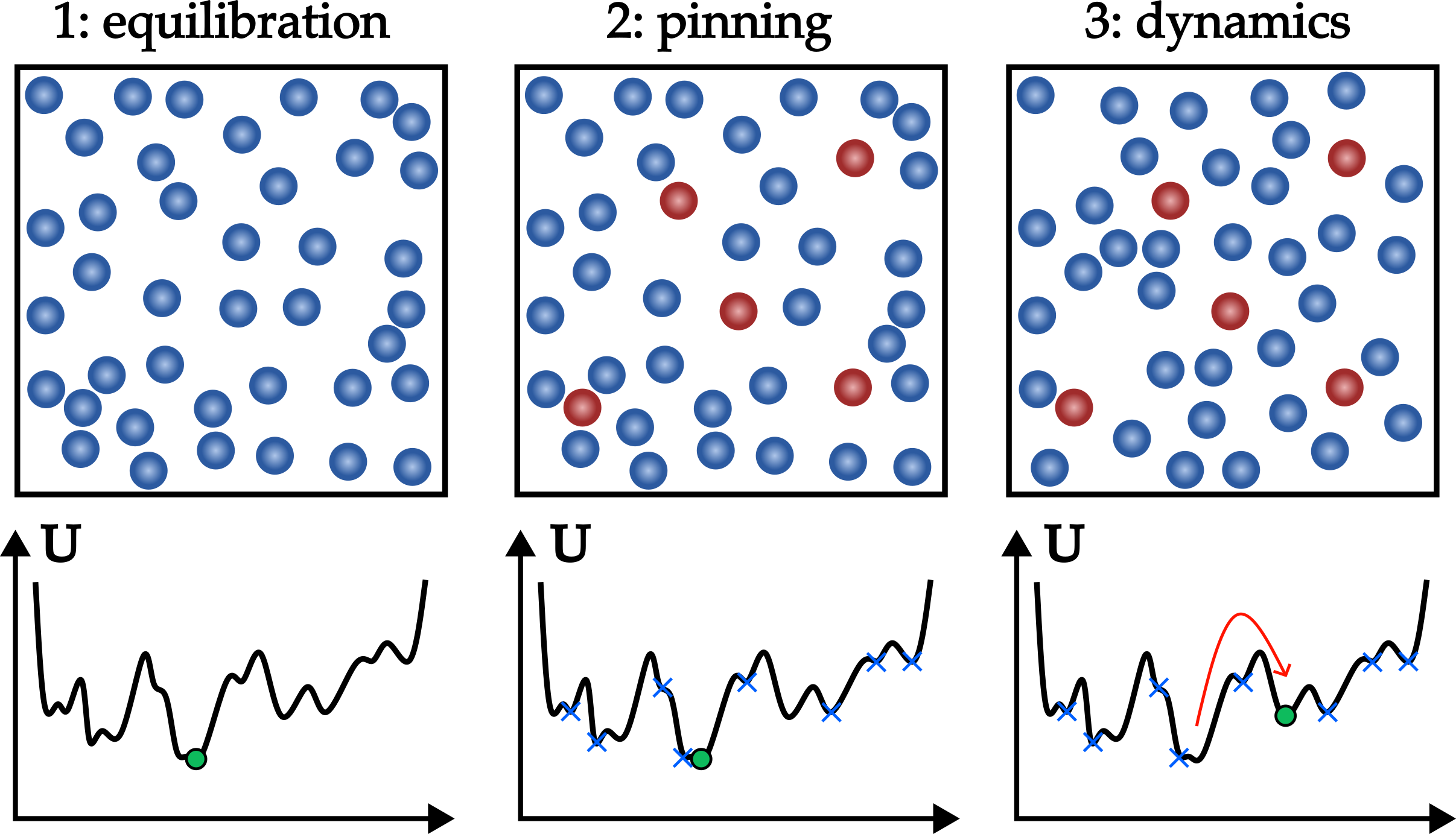}
    \caption{Starting from an equilibrium liquid configuration (top left), a fraction of particles is randomly pinned (shown in red, top middle), and the dynamics of the remaining mobile particles are subsequently examined (top right). Bottom: Corresponding schematic illustration of how random pinning modifies the potential-energy landscape. The accessible configuration space becomes restricted (indicated by blue crosses) upon the introduction of pinned particles, effectively increasing the energy barriers separating metastable states.}
    \label{fig:random_pinning}
\end{figure}

\vspace{0.25cm}
\noindent{\textbf{\textit{Algorithm}:}}
The random pinning method starts from an equilibrium bulk configuration of a glass-forming liquid at low temperature, irrespective of the specific interaction potential type. From this configuration, a fraction of particles are randomly selected and their positions are permanently frozen (or pinned)~\cite{kim2003,cammarota2012,kob2013,ozawa2015}, as schematically illustrated in Fig.~\ref{fig:random_pinning}. Alternatively, one can significantly increase the mass of the selected particles~\cite{anwar2024exploring}. We then assess the stability, for example by examining the enhancement of relaxation times of the remaining mobile (unpinned) particles in equilibrium~\cite{chakrabarty2015}, monitoring hysteresis in heating-cooling cycles~\cite{hocky2014}, and probing the mechanical response~\cite{bhowmik2019}. As one can imagine, due to the presence of pinned particles, the dynamics of the unpinned particles become progressively more glassy as the fraction of pinned particles increases. 
In this sense, random pinning provides an artificial way to solidify the system by introducing the concentration of pinned particles $c_{\rm pin}$ as an additional control parameter.

We note that the enhancement of stability induced by random pinning becomes more pronounced when the initial bulk configuration is taken at a lower temperature~\cite{kob2013,ozawa2015} 

It is also important to emphasize that if the pinned particles are selected from a dense, thermally equilibrated configuration, the remaining mobile particles remain in thermal equilibrium immediately after pinning~\cite{scheidler2004,krakoviack2010}, albeit with enhanced stability. 
In contrast, if the pinned particles are chosen e.g., in a purely random (Poisson) manner, 
thermal equilibrium is no longer guaranteed, and such an approach offers no algorithmic advantage.

\vspace{0.25cm}
\noindent{\textbf{\textit{Considerations}:}}
The enhanced stability induced by random pinning can be interpreted from the perspective of the potential energy landscape. 
Consider an initial bulk equilibrium configuration located at a local minimum of the energy landscape, corresponding to a relatively high-energy basin, as schematically illustrated in Fig.~\ref{fig:random_pinning}. 
Introducing pinned particles effectively restricts the accessible phase space of the system. Increasing the fraction of pinned particles progressively restricts the number of accessible metastable states available to the remaining mobile particles, thereby reducing the configurational entropy. As a result, the effective glass transition temperature shifts to higher temperatures (or lower densities), allowing the system to remain equilibrated deeper in the glassy regime. In this constrained energy landscape, configurations correspond to relatively deeper minima, which manifests as enhanced thermodynamic and kinetic stability and a pronounced slowdown of structural relaxation.

\vspace{0.25cm}
\noindent{\textbf{\textit{Applications}:}}
The random pinning protocol was first introduced by Kang Kim to study glassy dynamics by using the concentration of pinned particles as an additional control parameter, alongside temperature~\cite{kim2003}. 
This protocol has subsequently been generalized to other geometries, such as walls and cavities, in order to study the influence of pinned particles, including the extraction of associated correlation length scales~\cite{scheidler2004,biroli2008,yaida2016}. 
It was shown that thermal equilibrium for the unpinned particles is maintained when the pinned particles are selected from a dense, equilibrated bulk configuration~\cite{scheidler2004,krakoviack2010}. 
Cammarota and Biroli theoretically demonstrated that an equilibrium ideal glass (Kauzmann) transition can be achieved through random pinning~\cite{cammarota2012}, a prediction that was later validated by molecular dynamics simulations~\cite{kob2013,ozawa2015}. 
Thus, random pinning provides a unique approach to access an equilibrium glass state (where the configurational entropy vanishes) in computer simulations, without the need to explicitly solve the difficult equilibration or optimization problem.

Since then, various aspects of randomly pinned glass formers have been investigated, including their thermodynamics and dynamics in the supercooled liquid regime~\cite{charbonneau2013,jack2013,fullerton2014,chakrabarty2015,russo2015,nandi2022}, non-equilibrium heating-cooling cycles~\cite{hocky2014}, vibrational properties~\cite{ozawa2018b,shiraishi2022low}, and mechanical responses~\cite{bhowmik2019,mutneja2025}.

It is also worth mentioning that random pinning can be realized experimentally in colloidal glasses, where the positions of selected particles can be frozen using optical tweezers~\cite{gokhale2014,williams2018}.
For molecular systems, a similar idea has been explored by considering mixtures of two substances with different masses to mimic the pinning situation~\cite{kikumoto2020,das2023}.


\section{Random bonding}
\label{sec:random_b}

\noindent{\textbf{\textit{Description}:}}
Starting from a low-temperature configuration of a monomer (particle) system, pairs of particles are randomly selected and permanently bonded. 
The stability of the resulting molecular glass former is then investigated.

\vspace{0.25cm}
\noindent{\textbf{\textit{Algorithm}:}}
First, a low-temperature thermally equilibrium configuration of a particulate system is prepared (e.g., via standard molecular dynamics annealing). 
Starting from such a ``monomer'' configuration, one monomer is randomly selected, and a second monomer is chosen from its neighbors, typically defined as non-bonded particles located within the first coordination shell in the radial distribution function. 
A permanent bond is then introduced between this pair of monomers~\cite{ozawa2023}, forming a ``dimer.''
For bonding, one may employ either rigid-body constraints, such as those implemented in the RATTLE algorithm~\cite{andersen1983}, or harmonic potentials with sufficiently large spring constants.

\begin{figure}[t]
    \centering
    \includegraphics[width=0.99\linewidth]{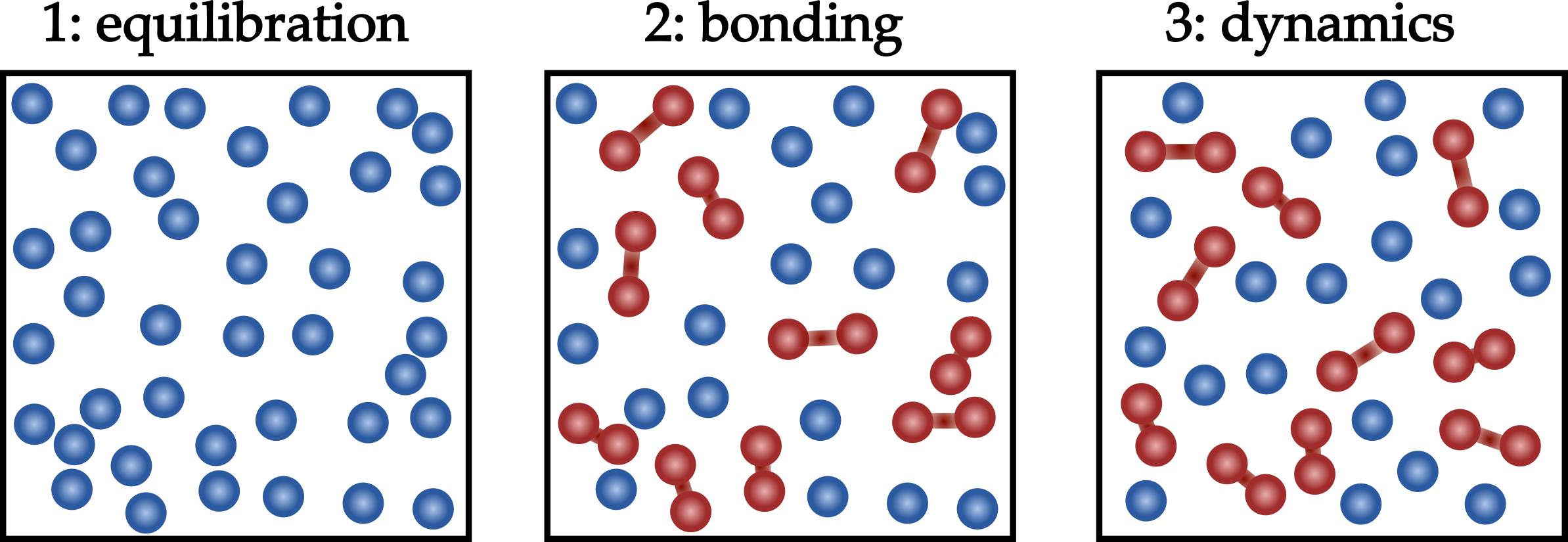}
    \caption{Starting from an equilibrium liquid configuration (left), a fraction of particle pairs is permanently bonded (middle), forming a stable molecular glass former. The resulting bonded system exhibits enhanced stability, characterized by significantly slower dynamics (right).
}
    \label{fig:random_bonding}
\end{figure}

This random pairing-and-bonding procedure is repeated until a desired fraction of dimers is obtained, which defines a control parameter analogous to the pinning concentration in random pinning protocols. 
As schematically shown in Fig.~\ref{fig:random_bonding}, this process yields a densely packed dimer configuration characterized by the dimer concentration $c_{\rm bond}$.

It is important to emphasize that the resulting stability is enhanced when the initial (monomer) configuration is prepared at a lower temperature and when $c_{\rm bond}$ is larger, akin to the behavior observed in random pinning~\cite{cammarota2012,kob2013,ozawa2015}. 
We also note that, although previous work has focused on randomly bonded dimers, the method can naturally be extended to trimers, polymers, or more complex molecular shapes.

\vspace{0.25cm}
\noindent{\textbf{\textit{Considerations}:}}
Random bonding is conceptually inspired by the random pinning method. 
The physical essence behind the success of random pinning lies in the introduction of constraints in phase space, which effectively narrows the accessible configurations (see Fig.~\ref{fig:random_pinning}). 
In the case of random bonding, the frozen degree of freedom is not the particle position itself but rather the interparticle distance between selected pairs, leading to the idea of bonding. 
In contrast to random pinning, random bonding does not break the translational invariance of the system, an important property of bulk materials. 
Consequently, this approach allows one to study mechanical responses such as brittle yielding and the emergence of sharp shear bands~\cite{ozawa2023}, phenomena that are typically absent in the yielding of randomly pinned glasses~\cite{bhowmik2019}.

Same as random pinning, it can be shown that if bonding pairs are chosen completely at random from all possible monomer pairs in the simulation box, the resulting molecular glass former remains mathematically guaranteed to be in thermal equilibrium immediately after bonding~\cite{ozawa2024}. 
However, if bonding pairs are chosen among neighboring monomers, which is physically more reasonable, this selection introduces a bias that formally violates the equilibrium guarantee. 
Nevertheless, numerical studies have demonstrated that the resulting configurations remain nearly equilibrated, showing only negligible deviations from equilibrium~\cite{ozawa2024}. 
Having said this, equilibrium guarantees are not a relevant issue when studying ultrastable glasses in non-equilibrium states.

\vspace{0.25cm}
\noindent{\textbf{\textit{Applications}:}}
The random bonding protocol was introduced in Ref.~\cite{ozawa2023}, where its stability was assessed through heating-cooling cycles, mechanical responses under shear deformation, and liquid-state dynamics. 
In particular, Ref.~\cite{ozawa2023} generated deeply annealed, nearly equilibrium configurations corresponding to relaxation times that are approximately $10^{7}$ times longer than those accessible by standard molecular dynamics simulations.
Moreover, non-equilibrium mechanical deformation tests revealed a pronounced stress overshoot, characterized by a stress-overshoot index $(\sigma_{\rm max}-\sigma_{\rm min})/\sigma_{\rm min}$ of approximately $1.4$, where $\sigma_{\rm max}$ denotes the peak stress and $\sigma_{\rm min}$ the steady-state stress after the overshoot.
Following this study, the issue of strict equilibration has been examined more thoroughly, both theoretically and numerically~\cite{ozawa2024}.
Besides, random bonding has been used to investigate how constraining the degrees of freedom modifies the fraction of unstable modes and its connection to glassy dynamics~\cite{sun2025}.

It was also suggested in Ref.~\cite{ozawa2023} that random bonding could be realized experimentally with existing techniques, for example by suddenly inducing attractive interactions between patchy colloids via changes in salt concentration, or by applying ultraviolet irradiation in colloidal systems with DNA linkers.

\section{Structural Optimization}
\label{sec:transient}
\vspace{0.25cm}

\noindent{\textbf{\textit{Description:}}}
The unprecedentedly deep equilibration enabled by the swap algorithm is largely due to the availability of unphysical moves i.e., the switching of particle sizes between particles. In a generalization of this approach, one can consider a situation where such variables were not only allowed to be swapped, but freely evolve. There are two ways in which this might be applied. In one, local particle sizes evolve in equilibrium with the chemical potential associated with a particular particle population~\cite{kapteijns2019} i.e., the definition of a equation of motion for particle size (grand canonical dynamics). The other is where unphysical moves might artificially optimize the system to have specific physical features which might be associated with ultrastability e.g., homogeneity of virial stress~\cite{leoni2025}. This second, non-equilibrium, approach is described below.

\begin{figure}[!t]
    \centering
    \includegraphics[width=0.99\linewidth]{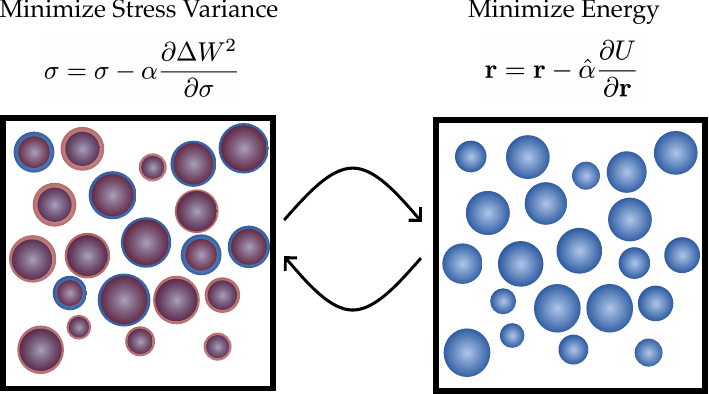}
    \caption{Structural optimization procedure combining stress homogenization and energy minimization. Starting from a glassy configuration, particle sizes $\sigma$ are first updated to reduce spatial variations in the local virial stress. The system is then re-equilibrated by minimizing the potential energy through updates of particle positions. These two steps are iterated until convergence, yielding configurations with reduced stress heterogeneity.}
    \label{fig:TrV}
\end{figure}

\vspace{0.25cm}
\noindent{\textbf{\textit{Algorithm}:}}
Starting from an initial, quenched glassy state, a local particle parameter is adjusted to achieve a target characteristic. A specific example might be in a Lennard-Jones system, when the target is a system with homogeneous local virial stress over space, as illustrated in Fig.~\ref{fig:TrV}. At each step, particles are individually resized by some value to reduce the variation in the virial stress over the particle population. The modified glassy state is then re-quenched to remove any force imbalances generated by the modifications. On comparing the new state to the original glass, the process may be repeated if the standard deviation in virial stress contributions is reduced. If not, the step size is likely too large, so needs to be scaled down. Convergence is arbitrarily defined, but may be reached when the standard deviation reaches some set value, or further reduction of particle size modifications does not yield a reduction in the standard deviation of the virial stress.

Regarding selection of the step size, there are many parallels with standard optimization techniques, such as steepest descent or conjugate gradient. The step size may be scaled by the susceptibility of the target parameter, much like step sizes are made smaller in an energy optimization when an energy minimum is approached.  For example, instead of resizing individual particles by a set, small amount (e.g. $0.01\sigma_i$, where $\sigma_i$ is the local Lennard-Jones size) to correct for locally low/high virial stress, one might scale the step size by considering how far the local virial stress is from the population average, and scale it by the gradient of the local virial stress as a function of particle size. In practice, this is further scaled by some factor (usually $<0.1$) to ensure that modifications are not too large such that unstable oscillations about a minimum are seen.

\vspace{0.25cm}
\noindent{\textbf{\textit{Considerations}:}}
The most significant limitation of these techniques is the change in the specific particle population of the initial glassy state. While it may be shown that the modified particle population quenched in a conventional manner gives rise to qualitatively different, rejuvenated glasses \cite{yanagishima2021,leoni2025}, it should be noted that properties which are specific to certain particle populations may be modified.

Another consideration is the nature of the steps themselves. While the steps in grand-canonical dynamics \cite{kapteijns2019} may be defined based on equilibrium with a set particle population, allowing for a form of detailed balance, optimization based algorithms are not constrained. Thus, the path taken during structural optimization as described here is strictly out of equilibrium, though it may remain close to it in some specific cases. This renders any analogies between optimization based methods and swap configurations non-trivial.

\vspace{0.25cm}
\noindent{\textbf{\textit{Applications}:}}
Early work by Brito, Wyart, and Lerner framed the swap algorithm as a special instance of grand-canonical dynamics \cite{brito2018,kapteijns2019}, where particle sizes may also be adjusted during equilibration within the constraint of a chemical potential defined particle size population with which a glassy state is in equilibrium. They noticed that relaxation of the strict particle population in this manner led to the generation of ultrastable glasses in significantly less time than what is required with slow-quenching protocols\cite{kapteijns2019}.

In parallel with this work, two independent strands of research led to the development of algorithms which leveraged modifications to particle size which were unconstrained by equilibrium with a particle reservoir. One was the study of glassy states with features of hyperuniform packings \cite{Torquato2018}. It was argued that amorphous hyperuniform packings might define a special instance of a well-equilibrated glass, a ``perfect'' glass \cite{zhang2016a}. To this end, an algorithm designed to create large, sparse hyperuniform packings \cite{Kim2019} was applied to create dense soft sphere packings \cite{Dale2022}. Unlike for sparse hard-spheres, modifications to soft spheres led to mechanically unstable states, necessitating mechanical relaxations and iterative adjustments until convergence was reached. The glasses generated in this work featured spatially correlated soft modes which broadly spanned the system size.

The correlation between such states and stability against glass aging was established through another body of work, where localized density inhomogeneities were correlated with the incidence of collective avalanchelike relaxation events in a deeply quenched glass \cite{Yanagishima2017}. In an effort to create glasses which were more stable to thermally activated relaxation, an algorithm inspired by \cite{Dale2022} was applied to modify quenched repulsive glasses to create glasses with a sharper distribution in local densities. The result was states that did not feature any of the avalanche relaxation events seen in the original glasses \cite{yanagishima2021}, and exhibited qualitatively different devitrification behavior when driven to crystallize through templating \cite{Yanagishima2023}.

While thermodynamic quantities such as bond-orientational order remained largely unchanged, it was found that the mechanical environment of individual particles was significantly modified to "homogenize" the response of the glass to perturbations. To achieve this more directly, an algorithm was proposed targeting not the local density, but the local virial stress. Importantly, homogenization of the local virial stress may be achieved in a much wider range of particulate systems. Applying the algorithm to binary glasses revealed a significant drop in energy and an enhanced kinetic stability which mirrored what is seen in physical vapor deposition glasses \cite{leoni2023,leoni2025}.

Target variables for this approach are not limited to density and virial stress. For example, optimization of local packing efficiency has given rise to ultrastable glasses with exceptionally low energy, as reviewed~\cite{berthier2026}. Tong {\it et al.} optimized a local packing parameter based on bond-angles \cite{Tong2019}, creating amorphous states which have many of the salient properties of a Debye solid \cite{fan2026}. Such properties were mirrored by density uniform packings with stronger convergence criteria \cite{wang2025} than in the original work \cite{yanagishima2021}. In work on a two-dimensional glass former, Bolton-Lum {\it et al.} \cite{boltonlum2026} modified local particle sizes to create strictly triangulated packings with a convergence in relaxation time with temperature following a Volger-Fulcher-Tammann (VFT) form approaching the ideal glass transition. These recent works exemplify efforts to explore the potential of target-oriented, non-equilibrium pathways to create ultrastable glasses.

Finally, the role of polydispersity remains less well understood than in swap-based approaches. Recent work has indeed shown that enhanced hyperuniformity and local ordering do not necessarily imply high thermodynamic stability~\cite{galliano2026}, concluding instead that the relaxation of particle-size degrees of freedom is the key ingredient for achieving ultrastability.

\section{Trajectory Sampling}
\label{sec:trajectory}

\noindent{\textbf{\textit{Description}:}}
Differently from schemes that alter the dynamics, trajectory sampling leaves the equations of motion unchanged while still giving access to long lived glassy states that are otherwise difficult to sample. Here we focus on the $s$-ensemble, where a Markov chain is performed in the space of trajectories, and where each path is assigned a statistical weight (the $s$ variable) that penalises dynamical activity~\cite{chandler2010}. In this way the method selects trajectories that are dynamically inactive and deeply metastable, even though such trajectories are rarely observed in equilibrium.

\begin{figure}[!t]
    \centering
    \includegraphics[width=0.99\linewidth]{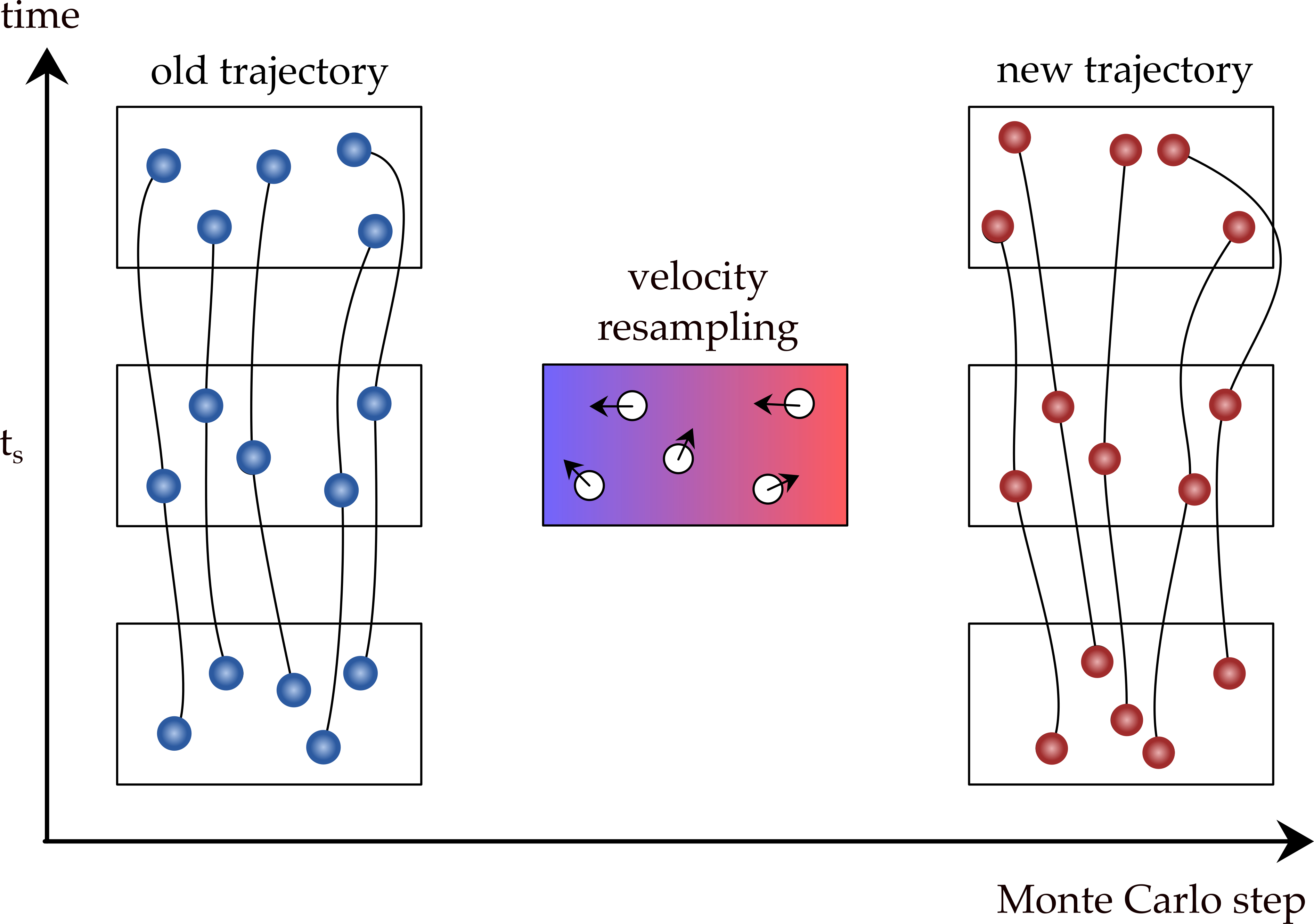}
    \caption{Schematic representation of trajectory sampling in the $s$-ensemble. Each column represents a full dynamical trajectory of duration $t_{\mathrm{obs}}$, with particle positions shown at a sequence of discrete physical times. A shooting move selects a configuration at an intermediate time slice (centre panel), perturbs the particle momenta, and regenerates the forward and backward parts of the trajectory to produce a new trial path. Repeating this procedure constructs a Markov chain in trajectory space, from which inactive and long lived trajectories are sampled with a weight that depends on their dynamical activity.}
    \label{fig:trajectory}
\end{figure}

\vspace{0.25cm}
\noindent{\textbf{\textit{Algorithm}:}}
The preparation of a biased trajectory in the $s$-ensemble proceeds as follows.
1) One starts from an initial trajectory of fixed duration $t_{\mathrm{obs}}$, generated by the unbiased dynamics (left panel in Fig.~\ref{fig:trajectory}). This initial trajectory (denoted as $X$) acts as the starting point for the Markov chain in trajectory space.
2) A time slice $t_s$ is chosen at random along the trajectory, and the configuration at that time is used as the shooting point.
3) The momenta at the shooting point are perturbed by a small random change, producing a slightly modified state from which new dynamics can be launched (middle panel of Fig.~\ref{fig:trajectory}).
4) From this state, the equations of motion are integrated forward and backward (by inverting momenta) to reconstruct a full trajectory of duration $t_{\mathrm{obs}}$. This yields a trial trajectory $X'$.
5) The dynamical observable associated with the $s$-ensemble is then evaluated along the trial trajectory. This observable is called the \emph{activity}, and measures the total amount of particle motion along the trajectory, for example through the sum of particle displacements between successive time intervals.
6) The trial trajectory is accepted with a probability that depends on the change in activity, using a Metropolis rule,
\[
p_{\mathrm{acc}} = \min\left\{1,\, e^{-s\,(C[X'] - C[X])}\right\},
\]
where $C[X]$ is the activity of trajectory $X$.
7) If the trial trajectory is accepted, it replaces the previous trajectory as the next state of the Markov chain; otherwise the previous trajectory is retained. Repeating this procedure generates a sequence of trajectories sampled from the biased ensemble.

\vspace{0.25cm}
\noindent{\textbf{\textit{Considerations}:}}
In the $s$-ensemble the biased distribution of trajectories is defined by the weight $\exp[-s\,C]$, where $C$ is the activity. The field $s$ acts as a conjugate variable to the activity, in close analogy with a Legendre transform: increasing $s$ suppresses particle motion and favours the sampling of slow, inactive trajectories. In this way the method continuously tunes the frequency of rare dynamical fluctuations that are essentially inaccessible in equilibrium sampling.

A central outcome is the identification of two distinct dynamical regimes~\cite{merolle2005,hedges2009,chandler2010}. For small values of $s$ the sampled trajectories remain liquid like and active, while for sufficiently large $s$ the ensemble becomes dominated by trajectories with strongly suppressed mobility. These regimes can be represented in proposed dynamical phase diagrams, where active and inactive trajectories occupy separate regions of trajectory space and are separated by sharp crossovers that become more pronounced at larger observation times. Such diagrams provide a useful framework for interpreting how rare dynamical fluctuations organise in space and time.

In addition to the standard shooting move, many other trajectory moves have been proposed within the transition path sampling framework~\cite{bolhuis2021}. These include shifting moves, which translate the whole trajectory forward or backward in time; partial path moves, which regenerate only a segment of the trajectory; and replica exchange in trajectory space, which facilitates sampling across different biasing fields or observation times. Such extensions can improve ergodicity and efficiency, especially when the dynamical landscape contains multiple competing inactive pathways or when the observation time is large.

A practical limitation of trajectory sampling is the restricted system sizes that can be treated. Each step of the Markov chain requires the generation and storage of an entire trajectory of duration $t_{\mathrm{obs}}$, so the computational cost grows quickly with both system size and observation time. As a result, existing studies focus on relatively small systems, of the order of a few hundred particles.


\vspace{0.25cm}
\noindent{\textbf{\textit{Applications}:}}
The conceptual foundation of trajectory based sampling originates from the development of transition path sampling, which introduced the idea that rare dynamical events can be studied by constructing a Markov chain in the space of trajectories~\cite{bolhuis2002}. This framework established shooting and shifting moves as practical tools for generating new dynamical paths without imposing a predefined reaction coordinate and demonstrated that entire trajectories can be treated as statistical objects.

Work on kinetically constrained models, analysed within dynamical large deviation theory, showed that rare fluctuations of activity can be accessed by introducing a field that biases trajectories according to their dynamical activity~\cite{ritort2003,jack2010}. This led to the formulation of the $s$-ensemble, in which the trajectory distribution displays distinct active and inactive regimes interpreted as a form of phase coexistence in space–time~\cite{merolle2005}. The same methodology was subsequently extended from simplified lattice models to atomistic glass formers, where analogous transitions between liquid like and inactive trajectories were identified~\cite{,hedges2009,chandler2010}.

A further development was the introduction of the $\mu$-ensemble, which biases trajectories according to their structural content, for example the time spent in locally favoured structures~\cite{speck2012,turci2017,turci2018,royall2020}. This approach established a direct link between dynamical large deviations and specific structural motifs associated with slow relaxation. For example, in Ref.~\cite{turci2017} the authors were able to equilibrate a Kob Andersen mixture down to $T\simeq 1.2 T_{\rm K}$, where $T_{\rm K}$ is the (estimated) Kauzamann temperature.


\section{Machine learning approaches}
\label{sec:ML}

The recent surge of machine learning (ML) techniques has revolutionized many areas of science, including computational physics and related fields. 
Consequently, there is growing anticipation that ML approaches will eventually tackle complex simulation problems in glass physics and even generate ultrastable glasses, much as ML has surpassed conventional or human-designed methods in numerous other domains. 
As of the time of writing this review, however, no ML-based methods are yet available that can generate ultrastable glasses in a way that significantly outperforms existing standard techniques. 
Nevertheless, we present a few pioneering works that may inspire future developments toward this ambitious goal.

Since ML-based approaches for glass generation are still in their infancy and no ready-to-use preparation protocols are currently available, this section departs from the structure used previously. Instead, we divide the discussion into three paragraphs, each presenting the essential information on the method and its main outcomes to date.

\subsection{Inverse design of ultrastability}


A novel strategy for generating mechanically stable glass configurations was proposed in Ref.~\cite{wang2021b}, which formulates the problem as an inverse-design task enabled by machine learning. In particular, Wang and Zhang~\cite{wang2021b} developed a Monte-Carlo-like update algorithm that iteratively modifies particle configurations so as to minimize the propensity for plastic activity under deformation, thereby steering the system toward increasingly mechanically stable glassy states.

Measuring plasticity for every Monte-Carlo trial configuration would be computationally prohibitive (typically quantified through the non-affine displacement measure $D_{\min}^2$~\cite{falk1998} under applied deformations). To overcome this bottleneck, the authors first trained a \emph{graph neural network} (GNN) to predict local plastic propensity directly from a static configuration, without explicitly performing deformation simulations. The training data for this surrogate model were generated from independent molecular dynamics simulations in which plastic events were computed athermally under a set of controlled deformations.
Once trained, the GNN serves as a fast and accurate surrogate for evaluating plasticity: it provides instantaneous predictions of local $D_{\min}^2$-like quantities, effectively replacing the need for repeated MD simulations. This surrogate model is then embedded within a Monte-Carlo loop, where the GNN-predicted plasticity plays the role of an ``effective Hamiltonian'' that defines the acceptance probability of configuration updates. 
In practice, the algorithm seeks configurations with progressively lower effective Hamiltonian (predicted plasticity) during the Monte-Carlo sampling.
The structural updates themselves are performed using swap Monte-Carlo, allowing the algorithm to explore configuration space far more efficiently than conventional particle displacements alone. Through this MC-like optimization process, the method progressively identifies configurations with reduced plastic propensity, yielding glass states that are more stable against mechanical deformation.

\begin{figure}[!t]
    \centering
    \includegraphics[width=0.99\linewidth]{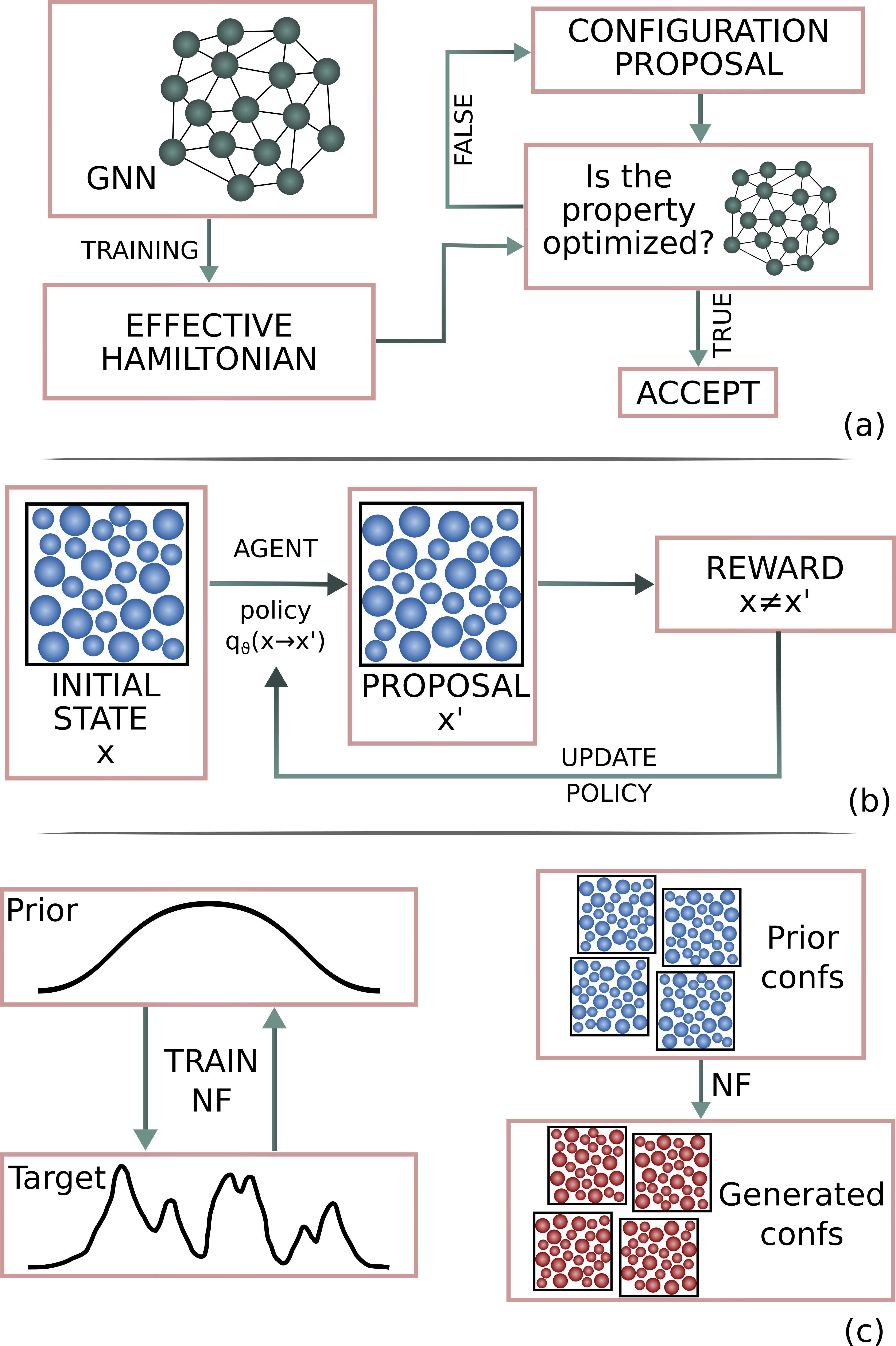}
    \caption{Three schematic examples of ML algorithms used to study supercooled liquids and glasses. (a) A Graph Neural Network is first trained to learn the mapping between configurations and one or more target properties, which define an effective Hamiltonian to be minimized. 
The trained GNN is then employed as a fast surrogate model that evaluates this effective Hamiltonian for any given configuration, and is subsequently used within a Monte-Carlo–like update scheme.
 (b) A reinforcement-learning-inspired strategy in which, starting from an initial state $x$, an agent generates a proposed configuration $x'$ according to a policy $q_\theta(x \to x')$. A reward is then computed, and the policy is updated accordingly. This procedure is repeated recursively. (c) A normalizing flow is trained to map a simple prior distribution onto a desired target distribution. 
After training, it generates configurations in the target distribution by applying this learned transformation to samples drawn from the prior.
}
    \label{fig:machine_learning}
\end{figure}

An overview of the algorithm is shown in Fig.~\ref{fig:machine_learning}(a). This method has been successfully applied to a binary Cu$_{64}$Zr$_{36}$ metallic glass model~\cite{wang2021b}, enabling targeted optimization of plastic propensity from MD deformation data. Notably, it reveals non-trivial glassy states, such as configurations that are geometrically stable yet energetically metastable-states that conventional thermal protocols cannot easily access. The study also demonstrates a degree of transferability across different alloy compositions.

Despite its promise, the approach has limitations. Its performance depends critically on the accuracy and generalization ability of the GNN, which may bias the sampling toward restricted regions of the configuration landscape, especially when separated by high energy barriers. Moreover, generating sufficiently diverse and labeled training data requires substantial computational resources.

Possible future developments include optimizing multiple target properties simultaneously, online retraining of the GNN as new configurations are discovered to improve predictive accuracy, incorporating reinforcement-learning-based or more advanced search strategies, extending the inverse-design protocol to other materials families via transfer learning, and improving scalability or integrating the method with experimental feedback loops. 

\subsection{Machine Learning assisted MC}

One promising application of ML techniques for generating stable glass configurations is ML-assisted Monte-Carlo sampling, where the proposal moves in the MC algorithm are optimized using a learning strategy.
Within the general formulation of Monte-Carlo framework~\cite{allen2017}, the acceptance probability $p_{\rm acc}$ for updating a configuration $x$ to a new configuration $x'$ is given by:
\begin{equation}
    p_{\rm acc} = \min \left\{ 1,\,
    \frac{q(x' \to x)}{q(x \to x')} \frac{\rho(x')}{\rho(x)}
    \right\} ,
    \label{eq:acceptance_probability}
\end{equation}
where $\rho(x)$ is the target Boltzmann distribution and $q(x \to x')$ is the proposal distribution for generating candidate configurations $x'$ from $x$. 
Equation~(\ref{eq:acceptance_probability}) is a generalization of the standard Metropolis algorithm, which corresponds to the case of symmetric proposals, $q(x' \to x)=q(x \to x')$.
Asymmetric updates introduced through a general proposal distribution $q(x  \to x')$ allow significantly greater flexibility in sampling configurations. 
Such flexibility has long been exploited through human-designed, system-specific insights to improve sampling efficiency~\cite{allen2017,frenkel2023}.
ML-assisted Monte-Carlo aims to take this idea further by constructing or learning an optimal proposal distribution $q_\theta(x  \to x')$, parameterized by a set of trainable parameters $\theta$, with the goal of accelerating sampling~\cite{liu2017}. 

While this approach has been used in various domains to tackle diverse problems, a recent and promising work implemented it specifically for investigating supercooled liquids~\cite{galliano2024}.
In this paper, by Galliano, Rende, and Coslovich~\cite{galliano2024}, the authors employed a policy-guided Monte-Carlo scheme inspired by reinforcement learning (RL)~\cite{bojesen2018}. 
In this approach, the proposal distribution $q_\theta(x \to x')$ (referred to as the \emph{policy} in RL terminology) is learned so as to maximize the magnitude of decorrelation (the \emph{reward}) induced by the Monte-Carlo update, such as particle displacements. 
The proposal distribution $q_\theta(x \to x')$ is parameterized in a physically motivated manner, combining displacement and swap moves, while also incorporating structural information such as local forces and local energies. A schematic illustration of the algorithm is provided in Fig.~\ref{fig:machine_learning}(b).
The hope is that ML can discover more flexible and effective proposal distributions compared to traditional (or human-designed) displacement or swap moves.

The authors reported an efficiency gain of roughly two orders of magnitude for soft-sphere models, whereas no significant improvement was observed for the Kob--Andersen mixture, indicating that the performance of the method is model-dependent.
Although the performance is still limited and therefore far from enabling the preparation of ultrastable glasses, the proposed framework is nevertheless promising. 
In particular, the design and parameterization of $q_\theta(x \to x')$ offer substantial room for improvement and further development, especially by incorporating collective moves using deep learning~\cite{bihani2023}.

\subsection{Sampling with generative models}

Generative models are making a huge impact across virtually all domains of science, engineering, and society. In essence, they learn or approximate an underlying probability distribution $\rho(x)$, which is generally unknown, from data by constructing a parametrized model $\rho_\theta(x)$ such that $\rho_\theta(x) \approx \rho(x)$. Using generative models to accelerate Monte-Carlo sampling has recently gained significant attention~\cite{noe2019,gabrie2022,marchand2023}. The core idea is that once the model $\rho_\theta(x)$ is learned, it can efficiently generate new configurations distributed according to $\rho_\theta(x)$, typically from Gaussian noise (or a high-temperature configuration). Leveraging this property, one can construct the proposal distribution in Eq.~(\ref{eq:acceptance_probability}) as
\begin{equation}
    q(x \to x') = \rho_\theta(x').
    \label{eq:proposal_generative_model}
\end{equation}
In other words, a new configuration $x'$ can be proposed directly from the generative model, without referring to the current configuration $x$ at all. Moreover, in the ideal limit where the generative model perfectly reproduces the target distribution, $\rho_\theta(x) = \rho(x)$, the Metropolis–Hastings acceptance probability becomes unity. In practice, the closer $\rho_\theta(x)$ is to $\rho(x)$, the higher the acceptance rate.
Thus, at each Monte-Carlo step, the simulation can propose nearly decorrelated configurations that are already very close to equilibrium.

In Ref.~\cite{jung2024}, Jung, Biroli, and Berthier employed a specific class of generative models, namely {\it normalizing flows} (NFs), for a ternary Lennard-Jones mixture. As schematically shown in Fig.~\ref{fig:machine_learning}(c), normalizing flows construct an explicit mapping between a simple {\it prior} distribution $\rho_{\rm P}(x)$, which is easy to sample (corresponding to high-temperature configurations), and an approximation $\rho_{\theta}(x)$ of more complex target distribution $\rho(x)$ of interest. Importantly, the target low-temperature distribution $\rho(x)$ does not need to be known or sampled in advance. Instead, the flow is trained iteratively using configurations drawn from an easier-to-sample reference distribution together with a reweighting scheme that incorporates the Boltzmann weight of the target temperature.

Once trained, the normalizing flow enables efficient sampling of configurations approximately distributed according to $\rho_\theta(x)$ by transforming samples drawn from the prior distribution $\rho_{\rm P}(x)$. These generated configurations can then be used as proposal moves according to Eq.~(\ref{eq:proposal_generative_model}), combined with a Metropolis-Hastings correction to ensure unbiased sampling of the target Boltzmann distribution.

Normalizing flows show strong potential for studying supercooled liquids, providing substantial speedups over conventional molecular dynamics and achieving performance competitive with advanced sampling techniques such as parallel tempering and population annealing, while still remaining inferior to swap Monte-Carlo. Their main current limitation lies in the accessible system size~\cite{jung2024}. Extending these approaches to larger system sizes is therefore an obvious and important next step.

A complementary line of work has developed global annealing procedures in which a generative model is trained sequentially at decreasing temperatures to learn the low-temperature distribution and coupled to local MC steps, thereby bridging generative methods and ML-assisted Monte Carlo sampling. This approach, however, has not yet been applied to structural glass formers~\cite{delbono2025}.

\vspace{0.25cm}

\section{Stability Across Methods}
\label{sec:stability}

\begin{table*}[!ht]
\centering
\begin{tabular}{|c||c||c||c|}
\hline
{\bf{Algorithm}} &{\bf{Kinetic stability}} &{\bf{Thermodynamic stability}} &{\bf{Mechanical stability}}\\
\hline\hline
Conventional  & $\mathcal{S} \simeq 7 \times 10^1$~{\setcitestyle{numbers,square}\cite{staley2015}}  & $\mathcal{R}_s\simeq 0.69$~{\setcitestyle{numbers,square}\cite{das2022b}}&  $\Delta\sigma/\sigma \simeq 0.5$~{\setcitestyle{numbers,square}\cite{richard2021}} \\
\hline
PVD  & $T_o/T_g \simeq 1.19$~{\setcitestyle{numbers,square}\cite{leoni2024}}, $\mathcal{S} \simeq 3 \times 10^2$~{\setcitestyle{numbers,square}\cite{leoni2024}} & --- & ---\\
\hline
Cyclic Shear & --- & --- & $\Delta\sigma/\sigma \simeq 0.9$~{\setcitestyle{numbers,square}\cite{priezjev2024}}  \\
\hline
\multirow{2}{*}{Swap} 
& \multirow{2}{*}{$\mathcal{S} \simeq 1.25 \times 10^4$~{\setcitestyle{numbers,square}\cite{fullerton2017}}} 
& $T_f/T_g\simeq 1.18$~{\setcitestyle{numbers,square}\cite{parmar2020b}}, $\mathcal{R}_s\simeq 0.2$~{\setcitestyle{numbers,square}\cite{berthier2019b}} 
& \multirow{2}{*}{$\Delta\sigma/\sigma \simeq 1.7$~{\setcitestyle{numbers,square}\cite{ozawa2018}}} \\
\cline{3-2}
& & $\tau_\alpha(\text{MC})/\tau_\alpha(\text{Swap})\simeq 10^{10}$~{\setcitestyle{numbers,square}\cite{ninarello2017}}  &  \\
\hline
Structural Optimization  & $T_o/T_g \simeq 1.39$~{\setcitestyle{numbers,square}\cite{leoni2025}} & --- & $\Delta\sigma/\sigma \simeq 3.0$~{\setcitestyle{numbers,square}\cite{kapteijns2019}} \\
\hline
Cluster Moves & --- &$\tau_\alpha(\text{Swap})/\tau_\alpha(\text{Chain})\simeq 10^2$~{\setcitestyle{numbers,square}\cite{ghimenti2024}} & --- \\
\hline
Random Pinning  & $\mathcal{S} \simeq 4.5 \times 10^2 $~{\setcitestyle{numbers,square}\cite{hocky2014}} & $\mathcal{R}_s\simeq 0 - 0.2$~{\setcitestyle{numbers,square}\cite{ozawa2015,ozawa2018b}} & $\Delta\sigma/\sigma \simeq 1.2$~{\setcitestyle{numbers,square}\cite{bhowmik2019}}\\
\hline
Random Bonding  & $T_o/T_g \simeq 1.75$~{\setcitestyle{numbers,square}\cite{ozawa2023}}& --- & $\Delta\sigma/\sigma \simeq 1.4$~{\setcitestyle{numbers,square}\cite{ozawa2023}} \\
\hline
Parallel Tempering & --- &  $\tau_\alpha(\text{MC})/\tau_\alpha(\text{PT})\simeq 10^{2}$~{\setcitestyle{numbers,square}\cite{flenner2006}}  & --- \\
\hline
Trajectory Sampling 
& $\mathcal{S} \simeq 10^1$~{\setcitestyle{numbers,square}\cite{speck2012}}
& $\mathcal{R}_s\simeq 0.43$~{\setcitestyle{numbers,square}\cite{turci2017}}, $T_{\text{K}}/T_{\text{eq}}\simeq 0.81$~{\setcitestyle{numbers,square}\cite{turci2017}}
& ---  \\
\hline
\end{tabular}
\caption{Comparison of stability in different glassy systems obtained with different methods. The symbols are defined as follows: $\mathcal{S}$ is the kinetic stability ratio. $T_o$, $T_g$, $T_f$, and $T_{\rm K}$ are respectively the onset, the glass transition, the fictive, and the Kauzmann temperatures. $\mathcal{R}_s$ is the configurational entropy ratio as defined in the main text. $\tau_\alpha$ is the relaxation time at the lowest accessible temperature for different methods. $\Delta\sigma/\sigma$ quantifies the stress overshoot at yielding. Some values were not explicitly reported in the original works and have therefore been inferred from the available data. 
}
\label{tab:1}
\end{table*}
1
In Sec.~\ref{sec:intro}, we outlined several approaches commonly used to characterize glass stability and presented a few representative quantitative indicators. We now systematically compare the computational strategies discussed thus far for producing ultrastable glasses by reviewing the stability metrics and quantitative results reported in the literature.
Because \emph{ultrastability} does not admit a unique definition and different algorithms emphasize distinct aspects of stability, direct comparisons are not straightforward. To address this issue, we organize the reported measures within a unified framework based on kinetic, thermodynamic, and mechanical properties of glasses.
In each case, stability is quantified through relative measures, typically by comparing properties of ultrastable glasses to those of conventional glasses or the equilibrium liquid branch. For clarity, representative quantitative indicators are summarized in Table~\ref{tab:1}, which enables a direct, though necessarily approximate, comparison between methods. While not exhaustive, this overview includes the most widely used and informative measures, providing a practical reference for assessing the relative performance of different algorithms in generating ultrastable glasses.


{\it Kinetic stability} is associated with the dynamical response and mobility of the glass system, particularly during reheating and melting processes, i.e., under non-equilibrium protocols.
A widely used measure of kinetic stability is the \emph{stability ratio} $\mathcal{S}$, defined phenomenologically and commonly employed in experiments. It is expressed as the ratio between the isothermal melting or transformation time $t_m$, i.e., the time required for a glass to transform into the supercooled liquid at a temperature $T>T_g$ (typically $T/T_g \simeq 1.25$), and the structural relaxation time $\tau_\alpha$ measured at the same temperature: $\mathcal{S} = t_m/\tau_\alpha$.
Values of $\mathcal{S}$ exceeding $10^3$-$10^4$ are generally considered characteristic of experimental ultrastable glasses. Values observed in simulations of ultrastable glasses prepared via physical vapor deposition (PVD) and random pinning are around $\mathcal{S} \simeq 10^3$, as reported~\cite{reid2016,leoni2024,hocky2014}. Swap Monte-Carlo simulations have produced even higher ratios, reaching $\mathcal{S} \simeq 10^4$ up to even $10^5$ in small systems~\cite{fullerton2017}, whereas trajectory-based sampling methods typically yield more modest values~\cite{speck2012}, $\mathcal{S} \simeq 10$.

A common phenomenological measure of kinetic stability compares the onset temperature $T_o$ of a glass, as previously introduced, to the glass transition temperature $T_g$, both determined at the same heating and cooling rates. 
For conventionally prepared glasses, this ratio depends on the applied cooling and heating rates, which affect both $T_g$ and $T_o$, although their values remain very close.
For ultrastable organic glasses prepared via vapor deposition, typical experimental values are $T_o/T_g \simeq 1.05$~\cite{tylinski2016,ediger2017}. 
In simulations, where rates are many orders of magnitude higher than in experiments, much larger ratios have been reported: $T_o/T_g \simeq 1.19$ for PVD glasses~\cite{leoni2024}, $T_o/T_g \simeq 1.39$ for virial-homogenized configurations 
~\cite{leoni2025}, and $T_o/T_g \simeq 1.75$ for random-bonding protocols~\cite{ozawa2023}. Indeed, in addition to increasing with stability, $T_o$ also increases with heating rate, preventing direct comparison of the ratio $T_o/T_g$ across different conditions.

An alternative approach to probe kinetic stability involves analyzing the melting dynamics through the Avrami exponent~\cite{fanfoni1998,herrero2023,chacko2024,leoni2025b} $n_A$, which characterizes the rate of phase transformation over time. It should be noted, however, that the exponent depends not only on the melting rate but also on factors such as the dimensionality of growing liquid clusters. While an Avrami exponent exceeding the standard value $d+1$ (with $d$ the spatial dimension) may indicate enhanced stability relative to conventional glasses, as found for a vapor deposited ice model with $d=3$ ($n_A\simeq4.4$ \cite{leoni2025b}) or for soft repulsive spheres in $d=2$ dimensions ($n_A \simeq 4.5$ \cite{herrero2023}), its value alone is insufficient to unambiguously determine the underlying transformation mechanism. Therefore, its interpretation must be supported by additional structural or dynamical evidence~\cite{leoni2025b}, and for this reason, it is not included in Table~\ref{tab:1}.


{\it Thermodynamic stability} is associated with fundamental quantities such as energy, enthalpy, entropy, as well as relaxation times, in {\it thermal equilibrium}. Several approaches have been proposed to quantify how deeply a glassy state lies within the (free) energy landscape. However, comparing energetic or temperature scales across different glass-forming systems is challenging due to their widely varying microscopic properties. In this context, entropy provides a more reliable basis for comparison, as it is directly related to the total number of accessible states at a given temperature.

In glassy systems, the configurational component of the entropy is of particular relevance, as discussed in Sec~\ref{sec:intro}. Moving in this direction, we consider the configurational entropy at the lowest accessible temperature (or highest density) and compare it to its value at the mode-coupling transition temperature, $T_{\mathrm{MCT}}$, when this quantity is available for the different methods, defining an associated stability ratio as $\mathcal{R}_s = S_c(T)/S_c(T_{\mathrm{MCT}})$. According to this criterion, recent simulations of conventional glass formers have been able to reach ratios of $\mathcal{R}_s \simeq 0.69$.~\cite{das2022b} Trajectory-sampling methods allow access to significantly deeper states, with estimated ratios around $0.43$,~\cite{turci2017} while swap Monte-Carlo simulations achieve even lower values, down to $\mathcal{R}_s \simeq 0.2$.~\cite{berthier2019b}
Random pinning allows one to access glassy states with very small values of the configurational entropy, $\mathcal{R}_s \simeq 0 - 0.2$, although the precise definition of the entropy in such constrained systems remains subtle~\cite{ozawa2015,ozawa2018b}.
A very recent study combining parallel tempering, swap Monte-Carlo, and population annealing in a carefully chosen order achieves $\mathcal{R}_s \simeq 0$ for relatively small system sizes~\cite{jung2025}.

Thermodynamic stability can also be assessed indirectly through relaxation times in thermal equilibrium. Although relaxation times are dynamical quantities, in the context of algorithms that employ unphysical moves they provide insight into how efficiently a given protocol can access low-energy states. For example, in swap Monte-Carlo simulations, the ratio of structural relaxation times between standard and swap dynamics at the respective lowest accessible temperatures has been conservatively estimated to be as large as $\tau_\alpha(\mathrm{MC})/\tau_\alpha(\mathrm{Swap}) \simeq 10^{10}$ in three-dimensions~\cite{ninarello2017}. Similarly, for irreversible chain-based collective-swap algorithms, a ratio of $\tau_\alpha(\mathrm{Swap})/\tau_\alpha(\mathrm{Chain}) \simeq 10^{2}$ has been reported at the highest accessible densities~\cite{ghimenti2024}, indicating a further enhancement in sampling efficiency beyond standard swap Monte-Carlo. By contrast, parallel tempering yields more modest improvements, with reported ratios of $\tau_\alpha(\mathrm{MC})/\tau_\alpha(\mathrm{PT}) \simeq 10^{2}$~\cite{flenner2006}.

In experiments, a way to assess thermodynamic stability is through the \emph{fictive temperature} $T_f$, defined as the temperature at which the glass structure would be in thermal equilibrium with the liquid \cite{ediger2017}. In simulations, however, the limited accessible time scales compared to experiments make the determination of $T_f$ less robust, and it is therefore infrequently evaluated. An exception is provided by swap Monte-Carlo simulations, where $T_f$ has been computed, showing that swap-generated configurations lie close to the experimental glass transition of the numerical model with $T_f/T_g\simeq 1.18$~\cite{parmar2020b}.

Alternative definitions of thermodynamic stability have also been employed in simulations. For example, in trajectory sampling of the Kob-Andersen mixture, thermodynamic stability can be evaluated using the ratio between (estimated) $T_\mathrm{K}$, the Kauzmann temperature, and $T_\mathrm{eq}$, the lowest temperature at which the inherent-structure energy remains well described by a quadratic approximation ($T_\mathrm{eq} \simeq 0.37$)~\cite{turci2017}.


{\it Mechanical stability}, which complements other measures of glass stability, refers to the material’s response to mechanical loading and can be quantified using rheological observables \cite{nicolas2018}. Under quasi-static deformation, glasses prepared with increasing stability exhibit a qualitative change in their stress–strain curves. In particular, the transition from conventional to ultrastable glasses is marked by a sharp, discontinuous change in the mechanical response \cite{ozawa2018,bhowmik2019,ozawa2023,wang2025}. The stress overshoot at yielding, $\Delta\sigma$, can therefore serve as an effective order parameter distinguishing these regimes. In Table~\ref{tab:1} we report $\Delta\sigma/\sigma=(\sigma_{\rm max}-\sigma_{\rm min})/\sigma_{\rm min}$, where $\sigma_{\rm max}$ and $\sigma_{\rm min}$ are the values of the maximum stress and its value just after the overshoot, respectively. The reported values range from the moderate value of $0.5$~\cite{richard2021} for conventional glasses to as high as $3.0$ for glasses prepared using structural optimization~\cite{kapteijns2019}.

Several alternative notions of stability have been proposed, often based on assumed correlations between specific structural or geometrical features and enhanced glass stability~\cite{kawasaki2014,jenkinson2017}. As these approaches do not provide direct quantitative measurements of stability, we do not include them in Table~\ref{tab:1}. Nevertheless, they remain relevant and will be collectively referred to here as structural stability.


The population of locally favored structures (LFS), i.e., geometric motifs that often locally minimize the energy, provides an efficient probe of structural stability. 
Several studies have shown that LFS populations correlate strongly with thermodynamic control parameters such as temperature or packing fraction~\cite{jenkinson2017}, and with dynamics~\cite{hocky2014b,jack2014,hallett2018,ishino2025,tanaka2025}. This correlation has been observed across a wide range of systems, including colloidal glasses~\cite{royall2008,leocmach2012,leoni2023} and metallic glasses~\cite{singh2013,royall2015}.
Icosahedra are found to be the most frequently observed LFS in many Lennard-Jones-like systems~\cite{frank1952,steinhardt1983,jonsson1988,jenkinson2017,leoni2023}. A larger concentration of icosahedra in UG with respect to CG has been found in glasses obtained with different methods, such as vapor deposition ~\cite{leoni2023}, structural optimization \cite{leoni2025}, or swap MC ~\cite{ozawa2017}.

More recently, machine-learning-based strategies have been developed to detect structurally relevant local environments in amorphous configurations. 
In both unsupervised~\cite{boattini2020} and supervised~\cite{martelli2020,leoni2021} approaches, high-dimensional order parameters are employed to identify local structural signatures associated with thermodynamic quantities such as energy~\cite{martelli2020,leoni2021}, or with dynamic properties such as dynamical heterogeneities~\cite{boattini2020,jung2025b}.

An alternative measure of glass stability is the centrosymmetry parameter, which quantifies the degree of local inversion symmetry in the arrangement of neighbors around each particle~\cite{milkus2016,liu2022,zaccone2026}. High centrosymmetry suppresses non-affine rearrangements that would otherwise reduce the shear modulus. In a perfectly centrosymmetric crystal, forces from opposite neighbors cancel under affine deformation, keeping particles in mechanical equilibrium. In disordered structures this cancellation is absent, generating residual forces that induce non-affine displacements.
Non-affinity is typically quantified by the parameter $\mu_{s,c}$, which measures how closely a glass mimics the elastic response of a crystal under shear or compression. Smaller values indicate a more ordered, crystal-like response \cite{tong2015, fan2026}, and thus a more stable glass.

Structural stability can also be explored through vibrational properties. In computer-generated glasses, the vibrational density of states, $D(\omega)$, universally follows $\omega^4$ scaling at low frequencies~\cite{buchenau1991,gurevich2003,richard2020,bonfanti2020}, with a prefactor that decreases as stability increases~\cite{wang2019}. Experiments on vapor-deposited glasses similarly show that low-frequency modes are progressively suppressed in more stable samples~\cite{swallen2007,perez2014,yu2015}. In the limit of extremely stable two-dimensional amorphous systems obtained through steric optimization, a crossover toward crystal-like Debye scaling $D(\omega)\sim\omega^{d-1}$ has recently been reported~\cite{fan2026}. The same study also identified broader signatures of emerging crystal-like order, including affine displacement and hyperuniformity, the latter manifested as suppressed long-wavelength density fluctuations with a small-$q$ spectrum $\chi(q)\sim q^{\alpha}$ approaching the crystalline limit $\alpha=d+1$ in $d$ dimensions~\cite{wang2025,fan2026}.

\section{Perspectives}
\label{sec:perspectives}

To help navigate this broad landscape, the algorithms discussed in this review can also be viewed through the lens of three encompassing paradigms, distinct from the five categories used to structure the presentation of individual methods, offering a complementary perspective on their underlying approach.

The first paradigm exploits the selective freezing or freeing of degrees of freedom. Swap Monte-Carlo exemplifies this logic: by temporarily liberating particle sizes as additional degrees of freedom, it enables rapid traversal of configuration space. Random pinning operates on the opposite principle, freezing a subset of particles to restrict phase space and lower configurational entropy. Physical vapor deposition fits naturally within this framework as well, since the enhanced mobility at the free surface provides faster degrees of freedom that allow local equilibration before particles are incorporated into the bulk. Structural optimization methods and random bonding follow analogous strategies, selectively activating or suppressing degrees of freedom either temporarily or permanently. Finally, cyclic shear can also be interpreted in these terms, with the strain amplitude acting as an effective control parameter that opens additional relaxation pathways in the energy landscape and grants access to deeper regions of that landscape.
Following a recent work \cite{ghimenti2026}, all these different preparation protocols, with the exception of the structural optimization, may access deeper states not by altering the underlying free-energy landscape, but by accelerating the exploration of configuration space effectively reparameterizing the flow of time along a common relaxation trajectory.

The second paradigm relies on biasing the target distribution and applying appropriate reweighting in order to effectively flatten free-energy or potential-energy barriers. This category naturally encompasses importance sampling methods, parallel tempering, population annealing, and trajectory-sampling techniques. The common principle is to enhance the sampling of configurations that would otherwise be exponentially suppressed in the equilibrium ensemble, while restoring statistical consistency with the Boltzmann distribution through reweighting. 

The third paradigm encompasses sampling methods that generate configurations through non-equilibrium dynamics with finite entropy production, while still ensuring that the stationary distribution follows the Boltzmann measure. Examples include lifted and irreversible Monte-Carlo schemes, where detailed balance is relaxed while global balance is maintained, enabling faster equilibration without affecting the equilibrium distribution.

Machine-learning-based approaches do not fit neatly into any single paradigm, but can in principle intersect with all three: ML models may implicitly learn to free or constrain effective degrees of freedom, construct biased proposal distributions for importance sampling, or generate non-equilibrium dynamics that enhance exploration while preserving the correct equilibrium measure. More broadly, machine learning and artificial intelligence have shown a remarkable ability to navigate high-dimensional spaces, as illustrated by their success in complex games such as Go~\cite{silver2018}, a capability that tends to emerge when the rules of the problem are well defined and performance is unambiguously measurable. Sampling a target distribution can be framed in precisely these terms: the objective is to decorrelate configurations as rapidly as possible while preserving the equilibrium measure. Although current ML-based approaches do not yet outperform state-of-the-art sampling algorithms, the preparation of ultrastable glasses is a well-posed computational problem, and the ability of ML to learn efficient exploration strategies in rugged, high-dimensional landscapes makes it a promising direction for future developments.


The quest for ultrastability remains however ongoing. From a theoretical perspective, it is intimately connected to the unresolved question of whether an ideal glass transition exists. At the same time, a pressing challenge is the extension of these sophisticated algorithms, often developed for simplified model systems, to more realistic glass formers featuring rotational degrees of freedom and more complex interaction potentials. Whether further progress will arise from entirely new algorithmic paradigms, from refinements of existing methods, or from hybrid combinations of current strategies remains an open question. 

A compelling recent example of combining different algorithms is provided by Ref.~\cite{jung2025}, which introduces a protocol merging parallel tempering, swap Monte-Carlo, and population annealing to achieve full equilibration of a two-dimensional Lennard-Jones ternary mixture down to zero temperature for systems up to $N=77$ particles and providing new insight into the nature of the Kauzmann transition in two dimensions.
The protocol first uses swap Monte-Carlo combined with parallel tempering to reach temperatures far below those accessible with standard approaches. At these low temperatures and for small system sizes, the potential-energy distribution develops a rare low-energy tail approaching the ground-state energy. Population annealing then amplifies these rare configurations through replication and reweighting, promoting them to typical equilibrium states. 
This work highlights how the careful combination and ordering of advanced sampling techniques can be crucial for achieving full equilibration, suggesting promising directions for future developments including larger system sizes, three-dimensional models and molecular liquids.

Taken together, the pursuit of ultrastable glasses will continue to drive methodological innovation, with advances that are likely to resonate well beyond the glass community, benefiting neighboring fields in statistical physics, soft matter, and computer science.

\begin{acknowledgements}
F.L. and J.R. acknowledge support by ICSC – Centro Nazionale di Ricerca in High Performance Computing, Big Data and Quantum Computing, funded by European Union – NextGenerationEU. 
M. O. thanks the support by MIAI@Grenoble Alpes and the Agence Nationale de la Recherche under France 2030 with the reference ANR-23-IACL-0006). 
T.Y. acknowledges support from JSPS KAKENHI Grant Number JP24K00597.
\end{acknowledgements}

\bibliography{biblio}


\end{document}